\documentclass[graybox]{svmult}

\usepackage{mathptmx}       
\usepackage{helvet}         
\usepackage{courier}        
\usepackage{type1cm}        
\usepackage{makeidx}         
\usepackage{graphicx}        
\usepackage{multicol}        
\usepackage[bottom]{footmisc}

\usepackage[dvipsnames]{xcolor}
\usepackage{graphicx}
\usepackage{natbib}
\usepackage{longtable}
\usepackage{tabularx}

\usepackage{ulem} 

\def\ltsima{$\; \buildrel < \over \sim \;$}
\def\simlt{\lower.5ex\hbox{\ltsima}}
\def\gtsima{$\; \buildrel > \over \sim \;$}
\def\simgt{\lower.5ex\hbox{\gtsima}}

\def\arcsec{$^{\prime\prime}$}

\makeindex             

\begin{document}

\title*{The origin of the stellar mass distribution and multiplicity}

\author{Yueh-Ning Lee, Stella~S.~R.~Offner, Patrick Hennebelle, Philippe Andr\'e, Hans Zinnecker, Javier Ballesteros-Paredes, Shu-ichiro Inutsuka, J.~M.~Diederik~Kruijssen}
\institute{Yueh-Ning Lee \at Department of Earth Sciences, National Taiwan Normal University, 88, Sec. 4, Ting-Chou Road, Taipei City 11677, Taiwan \email{ynlee@ntnu.edu.tw}
\and Stella S. R. Offner \at The University of Texas, Austin, TX 78712, USA. \email{soffner@astro.as.utexas.edu}
\and Patrick Hennebelle \at Laboratoire AIM, Paris-Saclay, CEA/IRFU/DAp -CNRS-Universit\'e Paris Diderot, 91191 Gif-sur-Yvette Cedex, France \email{patrick.hennebelle@cea.fr}
\and Philippe Andr\'e \at Laboratoire AIM, Paris-Saclay, CEA/IRFU/DAp -CNRS-Universit\'e Paris Diderot, 91191 Gif-sur-Yvette Cedex, France \email{philippe.andre@cea.fr}
\and Hans Zinnecker \at Universidad Autonoma de Chile, 
Nucleo de Astroquimica y Astrofisica, 
Avda Pedro de Valdivia 425, Providencia,
Santiago de Chile, Chile
 \email{hzinnecker50@gmail.com}
\and Javier Ballesteros-Paredes \at Instituto de Radioastronom\'\i a y Astrof\'\i sica, UNAM, Campus Morelia, Antigua Carretera a Patzcuaro 8701. 58090 Morelia, Michoacan, Mexico. \email{j.ballesteros@irya.unam.mx}
\and Shu-ichiro Inutsuka \at Department of Physics, Nagoya University, Chikusa-ku, Nagoya 464-8602, Japan \email{inutsuka@nagoya-u.jp}
\and J.~M.~Diederik Kruijssen \at
              Astronomisches Rechen-Institut,
              Zentrum f\"ur Astronomie der Universit\"at Heidelberg,
              M\"onchhofstra\ss{}e 12-14, 69120 Heidelberg, Germany, \email{kruijssen@uni-heidelberg.de}
}
%
%

\maketitle

\abstract{In this chapter, we review some historical understanding and recent advances on the Initial Mass Function (IMF) and the Core Mass Function (CMF), both in terms of observations and theories. We focus mostly on star formation in clustered environment since this is suggested by observations to be the dominant mode of star formation. The statistical properties and the fragmentation behaviour of turbulent gas is discussed, and we also discuss the formation of binaries and small multiple systems. }

\abstract*{
Each chapter should be preceded by an abstract (10--15 lines long) that summarizes the content. The abstract will appear \textit{online} at \url{www.SpringerLink.com} and be available with unrestricted access. This allows unregistered users to read the abstract as a teaser for the complete chapter. As a general rule the abstracts will not appear in the printed version of your book unless it is the style of your particular book or that of the series to which your book belongs.\newline\indent
Please use the 'starred' version of the new Springer \texttt{abstract} command for typesetting the text of the online abstracts (cf. source file of this chapter template \texttt{abstract}) and include them with the source files of your manuscript. Use the plain \texttt{abstract} command if the abstract is also to appear in the printed version of the book.}

\tableofcontents

\section{Introduction} \label{sec:intro}

Stars form in regions where a sufficient amount of mass is concentrated, reaching densities that allow gravitational collapse to overcome all possible supporting agents such as thermal and radiative pressure, turbulence, and magnetic fields. 
When such conditions are reached, the amount of mass usually allows a group of stars to form together, 
which is indeed supported by observations \citep{LadaLada03,Kruijssen12,Longmore+14}.   
Moreover, gravitational collapse is hierarchical and often leads to fragmentation across a range of size scales \citep{Efremov+98,hopkins2013a,VazquezSemadeni+17,pokhrel+2018,VazquezSemadeni+19}. 
Therefore, from the largest-scale entities that contain thousands to millions of star to small multiple or binary systems, 
there is always a certain degree of grouping, or clustering, when star formation takes place. 
In other words, stars that form in isolation represent a very small proportion.

Given a group of stars, statistical properties can be inferred, in particular that of the stellar mass, which is the primary parameter of a star.  A star's mass at birth, or ``initial stellar mass," determines how it will evolve, the way it interacts with its environment, how long it will live and the mechanism by which it will eventually die. 
The mass of members of a stellar cluster is of particular importance since it regulates the cluster evolution during the formation and also determines the disruption and survival of the cluster through stellar feedback and dynamics. 

This chapter focuses on the stellar Initial Mass Function (IMF), in particular, and the initial stellar multiplicity distribution, which is closely tied to the origin and measurement of the IMF. The IMF is of primordial importance in setting the formation, evolution, and survival of a cluster, as a consequence of the various  dynamical, thermal, and chemical regulation of the stellar activities coming from the whole range of the stellar mass spectrum. Good knowledge of the IMF is essential in many aspects. From an observational point of view, as will be discussed later, the IMF is not directly observable. Accurately translating observed luminosities to stellar masses requires understanding many of the fundamental properties of stellar evolution as well as the mechanisms governing cluster formation. On the other hand, a good parameterization of the IMF is crucial as it is often used as a sub-grid model in simulations of cluster or even galactic scales. Moreover, the IMF is important to many aspects of astronomy since stars are the basic building blocks of the Universe. 
Stars of mass around that of the Sun or lower are likely the most favorable for hosting planets and the existence of life.  
Their prevalence and interaction with companion stars are of primary interest for the search of life. On the other end of the spectrum,  massive stars are the most powerful engines inside clusters. They are the major producers of heavy elements that can influence the star formation of following generations. The shock waves or strong turbulence from their stellar feedback also dynamically affect the evolution of the host cluster and may even trigger the formation of nearby clusters. 

In this chapter, we will first discuss  stellar clusters as a whole. The characteristics of star-forming (and non-star-forming) cores will be described. We will then discuss the link between cores and stars. Within a stellar cluster, smaller bound, multiple systems of 2 or 3 stars are common, since about half of all stars have a stellar companion. We therefore dedicate a section to discussing stellar multiplicity. Finally, the effect of stellar feedback mechanisms on star-formation activity and the impact on the birth environment is also discussed. 

\section{Determination and universality of the Initial Mass Function}

\citet{Salpeter55} was the first to propose a functional description of the stellar mass spectrum at their birth. He compiled the stellar masses of field stars and found that the number in each mass bin follows a powerlaw distribution 
\begin{equation}\label{eq:IMF_dNdlogM}
\mathcal{N}(M) = dN/ d\log M \propto M^\Gamma
\end{equation}
where $M$ is the mass and $\Gamma = -1.35$ is now known as the Salpeter slope. 

Since then, observational advances have allowed the IMF to be derived for a variety of environments, including for a number of clusters, both in the solar neighborhood and in nearby galaxies. Integrated galactic IMF (IGIMF) are also measured for some galaxies. 
The stellar mass distribution exhibits a turnover mass around $0.2-0.3~M_\odot$ and decreases for lower mass stars, a characteristic which was not initially observed due to sensitivity limits. It was found that the observed mass spectra of stars seem to be strikingly similar in most of clusters \citep[see review by][and references therein]{Bastian+10}. 
Ever since, this mass function is referred to as the canonical stellar Initial Mass Function (IMF), which appears universal within the limits of observational uncertainties irrespective of conditions. Many functional forms have been proposed to describe the single star IMF, given all binaries are resolved; two of these are the most widely applied to present-day studies. 
The Kroupa IMF \citep[][and see a review of various analytical forms therein]{Kroupa02} is a piece-wise powerlaw function:
\begin{eqnarray}\label{eq:IMF_Kroupa}
\mathcal{N} = {dN \over d\log M} \propto \left\{ \begin{array}{lcl}  M^{0.7\pm 0.7} &{\rm ,} &0.01~M_\odot \le M < 0.08~M_\odot  \\ M^{-0.3 \pm 0.5} &{\rm ,}& 0.08~M_\odot \le M < 0.50~M_\odot\\ M^{-1.3 \pm 0.3}&{\rm ,}& 0.50~M_\odot \le M  \end{array} \right. .
\end{eqnarray}
The Chabrier IMF \citep{Chabrier05} connects a lognormal low-mass end to a powerlaw tail:
\begin{eqnarray}\label{eq:IMF_Chabrier}
\mathcal{N} = {dN \over d \log M} \propto \left\{ \begin{array}{lcl}  {\rm exp}\left[ -{ ({\rm log} M - {\rm log} 0.2)^2 \over 2 \times 0.55^2}\right] &{\rm ,}&M  \le 1~M_\odot \\ M^{-1.35} &{\rm ,}&M > 1~M_\odot \end{array} \right.
\end{eqnarray}
Both functions coincide with the Salpeter IMF for stellar masses above $\sim 1~M_\odot$.
However, as will be discussed in the following paragraphs, there are increasing evidences that the IMF might not be universal in some environments. The usage of these canonical functions thus should be taken with much care.

\subsection{Stars in clusters and associations: observational facts in the Solar neighborhood }
Direct study of the stellar mass distribution requires resolving individual stars in a group. 
Therefore, most of the available observation data, which span a significant range of stellar masses, come from nearby regions. 
The universality of the IMF is a conclusion mostly made from pioneering observations in the Solar neighborhood. 
In contrast, IMF universality has been challenged in some more extreme environments. 

While stars form mostly in groups, this does not necessarily mean that they stay in groups. 
Historically, clusters are classified into globular clusters (GCs) and open clusters (OCs), 
inside which the stars are observed to be coeval and have similar chemical composition most of the time.
GCs are the ancient and massive clusters associated with galactic spheroids, such as the bulge and halo. 
They are dense and contain $10^4-10^5$ stars within a few parsecs, making them strongly bound by gravity. 
OCs are found mostly in the galactic disk and are less massive.
However, more recent studies have indicated that there is no clear dichotomy between the two types of clusters.
Observations of young massive clusters (YMCs) are providing the missing links for how these structures form and evolve \citep[see reviews by][]{PortegiesZwart+10,Kruijssen14,Longmore+14,Bastian16}. 
YMCs are very massive clusters (typically  $> 10^4~ M_\odot$) with ages below $\sim 100$ million years. 
They have been suggested to be the modern counterpart of GCs, although exactly how GCs form remains debated. New theoretical models and galaxy formation simulations that apply physical descriptions for local-Universe cluster formation and evolution across cosmic history have recently successfully reproduced the properties of GCs by $z=0$ \citep[e.g.][and many others]{Elmegreen10,Kruijssen15,Pfeffer+18,Kruijssen+19,Li+19}. This adds credence to the idea that GCs may indeed be scaled-up versions of regular cluster formation.

Besides stellar clusters, which remain gravitationally bound from their birth to present day, 
some groups of stars that formed together are dispersed due to gas mass loss as a consequence of stellar feedback or perturbations from encounters with other clusters. Most field stars were not formed in isolation but instead were previously members of star clusters.
There are also entities called stellar associations, which typically contain hundreds of stars. These are the gravitationally unbound ``clusters'' of stars that have coherent velocities but do not have sufficient mass or density to keep themselves together. 
However, we can still reasonably infer that they formed together in the same molecular cloud environment and use them to study clustered star formation. In particular, such diffuse star-forming conditions may help to better understand the role of (the absence of) stellar interactions such as competitive accretion and stellar feedback. 
Indeed, some recent works suggest that associations are formed 
directly without ever forming a bound entity \citep[e.g.][]{Ward+18}. 

\subsection{Measuring the Initial Mass Function}

The stellar IMF is characterized by two major features: a high-mass slope of $dN/d\log M \propto M^{-1.35}$ \citep{Salpeter55} and a peak around $0.2-0.3 ~M_\odot$ . 
Given current observational limitations, it is not yet possible to clearly distinguish between the two most widely used analytical forms of the IMF at the low-mass end \citep[e.g., see review by][]{Offner+14}, i.e., a piece-wise power law or a lognormal. 
However, there are continuing efforts dedicated to constraining the low-mass end \citep[see e.g.][]{Downes+14,Muzic+15,Drass+16,Jose+17,ElBadry+17,Gennaro+18,Hoffmann+18,Megeath+19,Suarez+19,Muzic+19}. 
It is also worthwhile noting that  due to unresolved binaries, which are usually not corrected for in observations, the peak occurs at slightly higher masses than that described by eqs. (\ref{eq:IMF_Kroupa}) and (\ref{eq:IMF_Chabrier}).

Determining the IMF is a big observational challenge that comprises several steps. Since it is not possible to directly measure individual stellar masses, what is usually observed is the Present Day Luminosity Function (PDLF), which describes the number of stars in each luminosity bin. Models for the stellar physics are needed to convert the PDLF to the Present Day Mass Function (PDMF). However, several additional assumptions are required, because the stellar luminosity is a function of stellar mass, metallicity, age, and spin, which are difficult to measure and often poorly constrained. Stellar evolution models are needed to convert the PDMF to the IMF in order to correct for massive stars that might have evolved off the main sequence. Finally, dynamical models are required to correct for the evaporation of low-mass stars from the cluster and possible mass segregation for dynamically evolved clusters.  

When individual stars are not resolved, the IMF can still be measured using quantities integrated over the stellar population in the entire cluster or galaxy. These types of methods can not give strong constraints on the exact shape of the IMF but can still be used,  for example, to constrain the IMF slope \citep[see e.g.][for a summary]{Scalo1986book}. While these indirect methods introduce large uncertainties, they allow us to probe the possible IMF variation in a wider range of environments. We describe a few of the most commonly used techniques here. 

\begin{enumerate}
\item Mass-to-light ratio: Since stellar luminosity increases non-linearly with stellar mass, the ratio of the integrated mass and light values is sensitive to the IMF shape. However, in some cases it is debated whether excess mass comes from low-luminosity low-mass stars or from stellar remnants, i.e., high-mass stars that are no longer on the main sequence. 

\item Population synthesis: This is a forward modeling method. For a given stellar population, one can predict observables such as colors, spectra, or line intensities that can be measured and compared with those of a cluster or a galaxy. Finding an exact, or unique, solution is not always possible. However, this method is often used to test whether, for example, a canonical IMF is consistent with the observations. 

\item Chemical evolution models: Stars of different mass evolve on different tracks, and therefore the yields of various elements are functions of stellar mass. Measuring element abundances or isotopic ratios inside a cluster also constrains the shape of the IMF. This method, however, requires good knowledge of stellar evolution and carefully distinguishing between primordial enrichment and stellar production. Moreover, only the high-mass end of the IMF can be constrained using this technique.
\end{enumerate}

\subsection{The high mass end of the IMF: a power law}

The value of the high-mass slope, $\Gamma$, typically in the mass range from $\sim 1~M_\odot$ to several $10~M_\odot$, is the most easily measured from observations, since low-mass stars suffer from observational completeness and the most massive stars are statistically rare and short-lived. Therefore, this power law is often used to test the universality of the IMF and the many theories that aim to explain it.  The slope $\Gamma = -1.35$ is consistent within observational uncertainties for many observed regions in the Solar neighborhood \citep[see e.g.][]{Bastian+10}.
However, increasing number of observation results start to suggest that the IMF could be varying across different environments.
Table \ref{tab:IMF_slope} summarizes measurements for a variety of regions in the recent literature. 

Many of the measurements are actually presented in the form of the PDMF and the authors do not attempt to infer the IMF since extra uncertainties could be introduced. In young clusters that have not undergone much N-body relaxation or mass segregation, it is reasonable to assume that the PDMF is similar to the IMF. However, in some more evolved clusters \citep[e.g.][]{Habibi+13,Pang+13, Brandner+08,Bonatto+06}, shallower slopes are measured in the inner region -- probably because low-mass stars have been lost through evaporation or interaction with other clusters \citep[see e.g.,][for slope variation in globular clusters of different binding energy]{Kruijssen09,Paust+10}. Careful membership identification and dynamical modeling is required in these clusters to recover the IMF \citep[e.g.][]{Hosek+19}.

\begin{longtable}[t]{ p{0.4\textwidth} p{0.18\textwidth} p{0.2\textwidth} p{0.22\textwidth} } 
\label{tab:IMF_slope}
Region & $\Gamma$  & Mass range ($M_\odot$)  & description \\ \hline
{\bf Field} &&& \\
\citet{Mor+17} & $-3.2$ & $> 1$ & GD \\
\citet{Calamida+15} & $-1.41 \pm 0.50$ & $0.56-1.0$ & GB\\
& $-0.25\pm 0.19$ & $0.15-0.56$ & \\
\citet{Czekaj+14} & $-3.2$ & $> 1.53$ & GD \\
\citet{Zoccali+00} & $-0.33\pm 0.07 $ & $0.15-1.0$ & GB\\
\citet{Holtzman+98} & $-1.2$ & $0.7-1.0$ & GB \\
& $-0.3$ & $0.3-0.7$ & \\
\citet{Bochanski+10} & $-1.66\pm 0.10$ & $0.32-0.8$ & GD \\
& $0.02 \pm 0.15$ & $0.10-0.32$ & \\
\citet{Convey+08} & $-1.04$ & $0.32-0.7$ & GD \\
& $0.8$ & $0.1-0.32$ &\\
\citet{Reid+02} & $-0.3 \pm 0.2 3$ & $0.1-1.1$ & GD \\
& $-1.8 \pm 0.25$ & $0.11-0.3$ & \\
\citet{ReidGizis1997} & $[-0.5, 0.7]$  & $0.08-0.5$ & GD, 8 pc wide\\
& $-1.4 $  & $0.5-1.0$ & \\
\citet{Gould+97} & $-1.21$ & $0.59-1.0$ & GD \\
& $0.1$ & $0.08-0.59$ & \\
\hline
{\bf $\omega$ Centauri} \citet{Sollima+07} & $-1.3$ & $>0.5$ & Globular cluster\\
& $0.15$ & $0.15-0.5$ & \\
\hline
{\bf NGC 3603} &&& GD, starburst \\
\citet{Pang+13} & $-0.88 \pm 0.15$ & $1-100$ & Mass segregated \\
\citet{Harayama+08} & $-0.74_{-0.47}^{+0.62}$ & $0.4-20$ &  massive star-forming region \\
\citet{Stolte+06} & $-0.91 \pm 0.15$ & $0.4-20$ & \\
\citet{SungBessell04} & $-0.9 \pm 0.1$ & $1-100$ & flatter in inner region\\
\hline
{\bf Westerlund 1} &&& Galactic disk \\
\citet{Andersen+17} & $-1.32\pm 0.06 $ & $0.6-1.4$&\\
& $-0.25 \pm 0.10$ & $0.15-0.6$  & \\
\citet{Lim+13} & $-0.8 \pm 0.1$ & $5 -100$ & \\
\citet{Gennaro+11} & $-1.44^{+0.08}_{-0.20}$ & $3.5-27$ & \\
\citet{Brandner+08} & $-0.6$ & $3.4-27$ ($<0.75$ pc) & \\
& $-1.3$ & ($0.75-1.5$pc) & \\
\hline
{\bf Westerlund 2} &&& Galactic disk \\
\citet{Zeidler+17} & $-1.46\pm 0.06$ & $0.8-25$& PDMF\\
\citet{Ascenso+07} & $-1.20 \pm 0.16$ &  $>0.8$ & \\
\hline
{\bf Trumpler 14 \& 16} \citet{Hur+12} &$-1.3 \pm 0.1$& $>1.6$ & GD\\
\hline
{\bf h \& $\chi $ Persei} \citet{Slesnick+02} & $-1.3 \pm 0.2$ & $4-16$ & GD PDMF \\
\hline 
{\bf Arches} &&& CMZ YMC \\
\citet{Hosek+19} & $-0.76 \pm 0.08$ & $>1.8$ & IMF \\
\citet{Habibi+13} & $-0.5 \pm  0.35$ & ($<0.2$ pc)&  PDMF, mass segregated\\
& $-1.21 \pm  0.27$ & ($0.2-0.4$ pc)&  \\
& $-2.21 \pm  0.30$ & ($0.4-1.5$ pc)&  \\
\citet{Espinoza+09} & $-1.1\pm 0.20$ & $10-100$ &  PDMF, flattening toward the center\\
\citet{Kim+06} & $-0.91 \pm 0.08 $ & $1.3-50$& inner region PDMF \\
\citet{Stolte+02, Stolte+05} & $-0.8 \pm 0.2$ & $6-65$ & inner region PDMF\\
\hline
{\bf Young Nuclear Cluster (YNC)} &&& GC  YMC\\
\citet{Lu+13} & $-0.7 \pm 0.2$ & $> 10$ &  \\
\hline
{\bf Quintuplet} \citet{Hussmann+12} & $-0.68^{+0.13}_{-0.19}$ & 5-40 &  YMC CMZ PDMF \\
\hline
{\bf M31} \citet{Weisz+15} & $-1.45^{+0.06}_{-0.03}$ & $> 1$ & 85 young clusters \\
\hline
{\bf NGC 6231} 
\citet{Sung+13} & $-1.1\pm 0.1$ & $0.8-45$ & YOC \\
\hline
{\bf NGC 2264} 
\citet{SungBessell10} & $-1.7\pm 0.1$ & $>3$ & YOC \\
\hline
{\bf NGC 6611} \citet{Bonatto+06} & $-1.45\pm 0.12 $ & $5-25$ & YOC, mass segregated\\
& $-1.52\pm 0.13 $ & (Halo) & \\
& $-0.62\pm 0.16 $ & (Core) & \\
\hline
{\bf LMC} \citet{Gouliermis+06} & $-2.1 \pm 0.1$ & $0.73-1.03$ & \\
& $-4.6 \pm 0.1$ & $1.0-2.4$ &  Steeper than Salpeter \\
\hline
{\bf R136} \citet{Andersen+09} & $-1.2\pm0.2$ & $1.1-20$& LMC starburst 30 Dor (NGC 2070)\\
\hline
{\bf Lindsay 1} \citet{Glatt+11} & $-0.51 \pm 0.11$ & 0.63-0.93 & SMC PDMF \\
\hline
{\bf NGC 339} \citet{Glatt+11} & $-1.29 \pm 0.15$ & 0.56-0.97 & SMC PDMF \\
\hline
{\bf Lindsay 38} \citet{Glatt+11} & $-0.74 \pm 0.17$ & 0.57-0.94 & SMC PDMF \\
\hline
\caption{Measured IMF slopes from the literature. There is a very nice compilation of earlier observations by \citet{Kroupa02}, thus we selectively show more recent results. Note that many of the measurement are the PDMF, of which the authors have not made an attempt to recover the IMF. This list is not meant to be exhaustive but to give a flavor of possible environmental influences on the IMF slope. (GB: Galactic bulge; GD: Galactic disk; GC: Galactic Center; SSC: Super Star Cluster; CMZ: Central Molecular Zone; YOC: Young Open Cluster; LMC: Large Magellanic Cloud; SMC: Small Magellanic Cloud)}
\end{longtable}

Stars within a given cluster are coeval most of the time. 
The age spread, which is generally small ($<1$~Myr) with respect to the cluster age \citep[e.g.][]{Longmore+14}, can be used to infer the duration of active star formation inside the cluster. 
From previous chapters (Ch.~6 and Ch.~7), we have already seen that molecular cloud exhibit a wide range of masses and sizes, leading to a large variety of star-forming environments. 
The question then becomes: what makes the IMF so universal? 
What physical mechanisms actually regulate the star formation at stellar scales?
Is the IMF universal in environments that deviate significantly from the Solar Neighborhood? 

Recent observations give some hints about possible deviations from a universal IMF in extreme environments. 
For example, \citet{Hosek+19} resolved the Arches cluster in the Central Molecular Zone (CMZ; the central few 100~pc of the Milky Way) and found the IMF to be top-heavy, i.e., abundant in massive stars. Early-type galaxies (ETGs) show an excess mass-to-light ratio with respect to the IMF. 
This indicates either abundant brown dwarfs or massive stars that have already reached the end of their lives \citep{vanDokkumConroy10,vanDokkumConroy12,Cappellari+12}.
Using $^{13}$CO and C$^{18}$O line intensity ratios, \citet{Zhang+18} suggested that the IMF is more top-heavy in actively star-forming galaxies, in particular within starburst galaxies. There has also been some evidence showing that the IMF may be more top-heavy in high-density and low-metallicity environments \citep{Marks+12}.

In terms of theory and simulations, there are some existing studies of the extreme environments. Following the same reasoning of the gravo-turbulent model of molecular cloud fragmentation \citep{HennebelleChabrier09}, \citet{Chabrier+14} proposed that the dense and turbulent environment of ETGs should lead to a bottom-heavy IMF that peaks at lower mass. 
Most of the existing numerical studies actually show a power law mass spectrum of $\Gamma \sim -1$, actually shallower than the Salpeter value (see \S~\ref{sec:IMF_slope_reservoir} and \S~\ref{sec:IMF_simu}). 
Recently, \citet{lee2018a} showed that in a globally collapsing cloud dominated by turbulent support, the mass function likely follows $\Gamma \sim -0.75$, 
while the spectrum becomes flat with $\Gamma \sim 0$ if the thermal pressure is important. 

The latter case is compatible with the numerical finding that primordial clusters likely have a flat mass spectrum. 
Many authors have suggested that, due to the deficiency of molecular line cooling, high temperatures cause population III stars to be more massive such that their IMF peaks at a much larger mass than the canonical IMF \citep[e.g.][]{UmedaNomoto02,NakamuraUmemura99,OmukaiNishi99,Bromm+99}.
Numerical simulations of population III star formation in the primordial universe \citep[e.g.][]{HiranoVolker17, Clark+11, Greif+11} indeed suggest that first a primordial mini-halo forms a massive star surrounded by a disk, and subsequent stars are formed from disk fragmentation. This mode of star formation likely gives a top-heavy IMF dominated by one massive star at the center of the cluster. However, observational confirmation of the population III IMF remains a big challenge.

Overall, there are increasing hints for possible IMF variation, particularly in extreme environments, while \citet{Parravano+2018} also caution the possible bias from limited sample volume. Efforts are continuing in both observational and theoretical work to determine the key physics that sets the IMF and define the conditions for which it possibly varies.

\section{Dense core properties}\label{sec:core_obs}

\subsection{Observations of various core types and their definitions }\label{sec:core_obs_difinition}

Conceptually, a {\it dense core} may be defined as a molecular cloud fragment, locally denser than its surroundings, that can potentially form an individual star (or a small multiple
system) by gravitational collapse. 
Dense cores are thus the smallest molecular cloud units within which star formation can occur \citep[e.g.][]{Bergin+07}.   
This is in contrast to massive {\it clumps} out of which star clusters form within giant molecular clouds (GMCs) \citep[e.g.][]{Williams+00}. 
These definitions 
correspond to density structures that are now routinely observed in Galactic clouds thanks to submillimeter and millimeter maps 
in both molecular lines and dust continuum emission. 
Observationally, it is however not always straightforward to decide whether a given cloud fragment will form a single star, multiple stars, 
or no stars at all. 
Strictly speaking, 
the classification of structures identified in observational surveys is therefore always tentative to some extent. 
With the advent of powerful infrared and submillimeter facilities, such as {\it Spitzer}, {\it Herschel}, IRAM, and ALMA, this classification 
has nevertheless become more and more secure.
Historically, the first examples of dense cores were detected as dark globules in dust extinction \citep{Bok+47} 
or as blobs of dense molecular gas in line transitions of NH$_3$  \citep[e.g.][]{Myers+83}. 
The first survey for dense cores in nearby molecular clouds was performed in NH$_3$ \citep{Benson+89}.
Nowadays, the most extensive and sensitive surveys for dense cores are carried out in optically thin 
submillimeter continuum emission from cold dust 
\citep[e.g.][]{Ward-Thompson+94,Motte+98,Johnstone+00, kony2015, HKirk+16}, 
as this technique can deliver column density maps of molecular clouds with the highest dynamic range (both in terms 
of density and spatial scales), especially when used from space with, e.g., {\it Herschel}. 

\setlength{\unitlength}{1cm}
\begin{figure*}
\begin{picture} (0,6)
 \put(0.5,0){\includegraphics[width=11cm,angle=0]{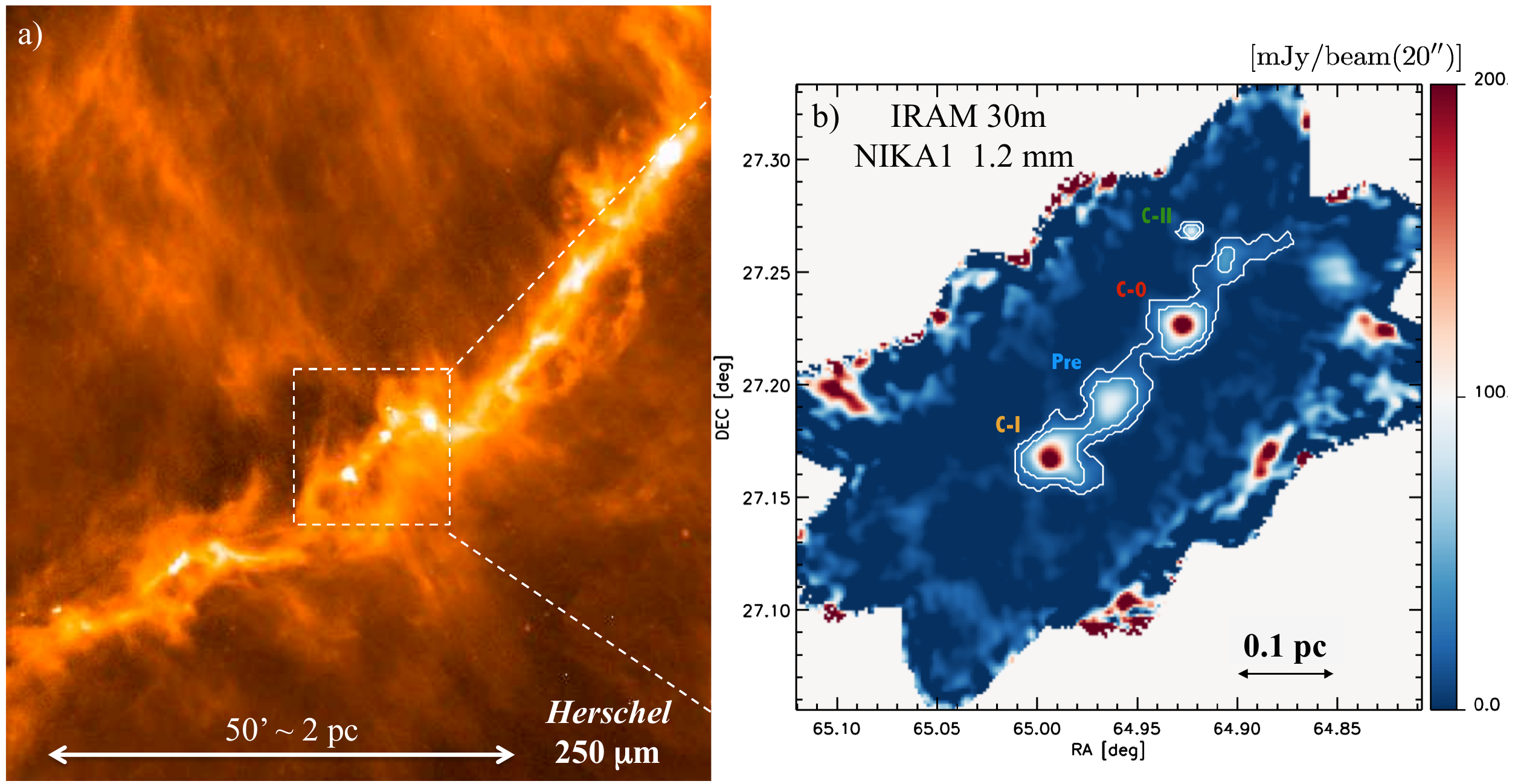} }
\end{picture}
 \caption{
 (a) {\it Herschel}/SPIRE 250~$\mu$m dust continuum image of the Taurus B211/B213 filament in the Taurus molecular cloud 
 \citep{Palmeirim+13,Marsh+16}. Note the presence of several prominent dense cores along the filament. 
(b) IRAM/NIKA1 1.2~mm dust continuum image of the central part of the {\it Herschel} field on the left
(from \citealp{Bracco+2017}), 
showing a chain of at least three prominent and quasi periodically-spaced dense cores, including a prestellar core (at the center) 
and two (Class~0/I) protostellar cores.}
\label{taurus_cores}
\end{figure*}

In practice, the immediate vicinity of each local maximum in column density maps 
derived from submillimeter continuum imaging data (cf. Fig.~\ref{taurus_cores}) 
is identified as a (candidate) dense core. 
One may also use significant breaks in the gradient of the column density distribution around each 
core peak to define the core boundaries (cf. Fig.~\ref{core_profiles} and \citealp{Roy+14}).
This can be a difficult task, however, unless the core is relatively isolated 
and the instrumental noise in the data is negligible (as is fortunately the case with {\it Herschel} observations of nearby clouds).
Ideally, maps that resolve the local Jeans length and the sonic scale 
are required to make sure that the candidate core 
does not have significant substructure and, e.g., is not on its way to fragmenting into a small star cluster. 
In dense ($A_V \geq 10$) parts of molecular clouds, this typically requires a spatial resolution 
significantly better than $< 0.05\,$pc for gas at $\sim \,$10\,K, corresponding to an angular resolution 
significantly better than 30\arcsec --60\arcsec  in nearby regions at $d\sim \,$140--300\,pc. 
For reference, {\it Herschel} data have a resolution of 18\arcsec ~at $\lambda = 250\, \mu $m.

Dense cores can be divided into several categories. 
A starless core is a dense core with no associated protostellar object
and may be gravitationally bound or unbound. 
A prestellar core is a dense core that is both starless and self-gravitating. 
A protostellar core is a dense core within which a protostar has already formed, 
where evidence of the latter comes from the detection of an embedded infrared source \citep[e.g.][]{Beichman+86},
a compact radio continuum source, or a bipolar molecular outflow \citep[e.g.][]{Andre+93}. 

Starless dense cores are observed at the bottom of the hierarchy of interstellar cloud structures and depart from 
the \citet{larson1981}
self-similar scaling relations (see \citealp{Heyer+09} 
and \S~4.5 in Chap.~7 
for a generalized version of the Larson relations).  
In particular, starless cores are characterized by subsonic levels of internal turbulence and may be described as  
islands of quiescence embedded in a sea of supersonically turbulent gas 
corresponding to their parent molecular cloud \citep{Myers1983,Goodman+98,Caselli+02,Andre+07,Pineda+2010}.
In other words, while starless dense cores are associated with local peaks in column density maps, 
they correspond to local minima in maps of the gas velocity dispersion \citep[e.g.,][]{Friesen+17,Chen+2019a}.
Observationally, it is remarkable that all cloud structures with supersonic line-of-sight velocity dispersions tend 
to be highly fragmented into significant substructures when observed at sufficient resolution, while structures with transonic 
or subsonic velocity dispersions show little substructure (see, e.g., Fig.~6 of \citealp{ward-thompson2007} and \citealp{Dunham+16,Kirk+2017}). 
Starless dense cores, which by definition are single fragmentation units, are therefore expected to 
have diameters below 
the sonic scale, or typically $\simlt 0.1\,$pc in low-density molecular gas and sometimes $\ll 0.1\,$pc in high-density environments.

\setlength{\unitlength}{1cm}
\begin{figure*}
\begin{picture} (0,6)
 \put(0.25,0){\includegraphics[width=11cm,angle=0]{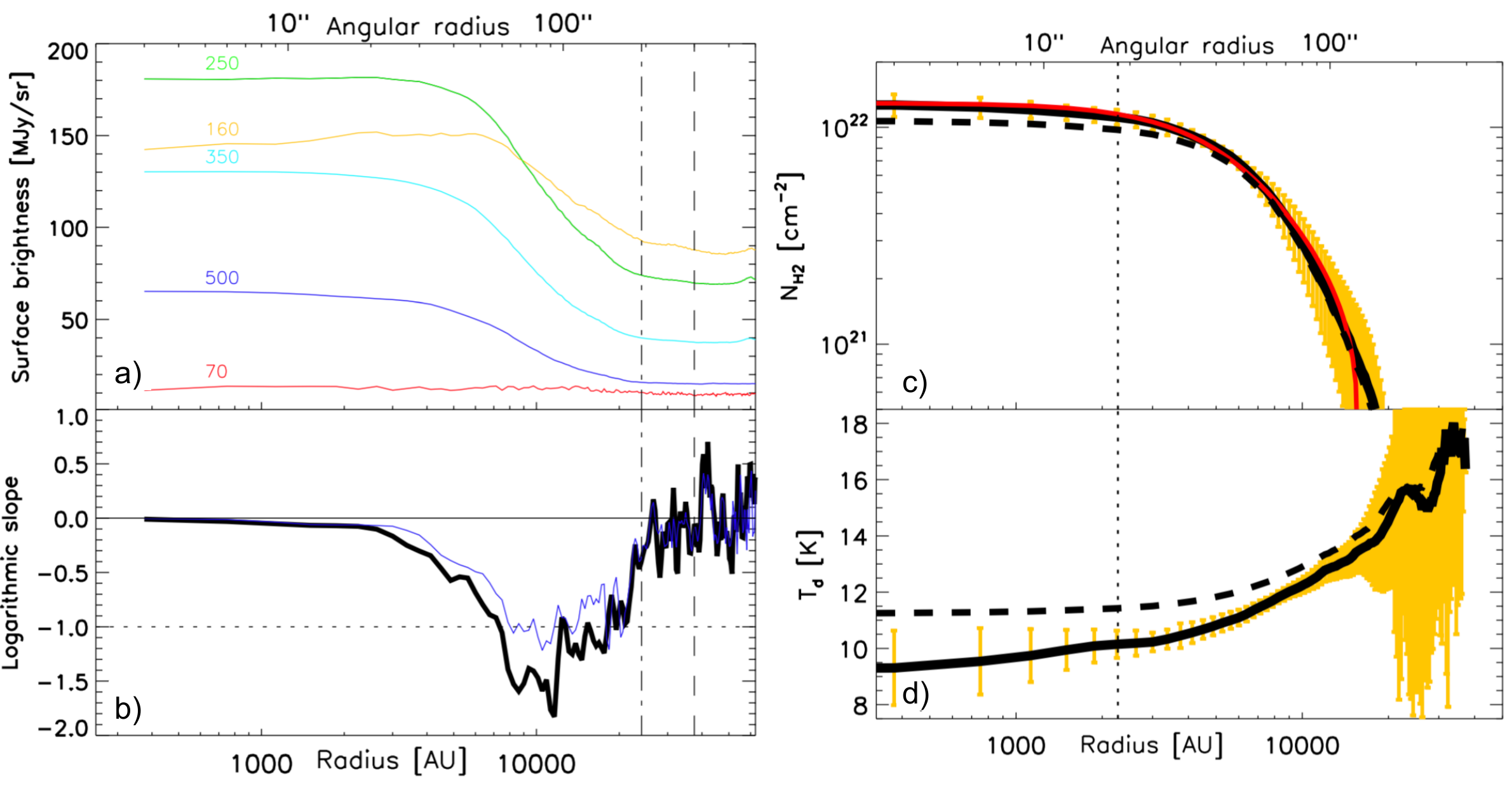} }
\end{picture}
 \caption{
(a) Examples of circularly-averaged radial intensity profiles derived from {\it Herschel} data for a prestellar dense core 
at 70, 160, 250, 350, 500~$\mu$m.
(b) Logarithmic slopes of the 
column density profile (black solid curve) and 500~$\mu$m intensity
profile (blue solid curve) as a function of radius, for the same core.
The dotted horizontal line is the logarithmic slope $s \equiv d\,{\rm ln}\, {N_{\rm H_2}}/d\,{\rm ln}\,r  = -1$ 
expected for a $\rho \propto r^{-2}$ density profile. 
Note how the core boundary can be reasonably well defined as the first point away 
from core center where the logarithmic slope goes back to $s \sim 0$ 
and the contribution from the core merges with the parent cloud emission.  
(c) Radial column density profile derived from the same {\it Herschel} data. 
(d) Dust temperature as a function of radius for the same pretsellar core. 
(Adapted from \citealp{Roy+14}.)}
 \label{core_profiles}
\end{figure*}

To first order, known prestellar cores have simple, convex (not very elongated) shapes, and have flat-topped 
radial density profiles approaching the density structure of Bonnor-Ebert (BE) isothermal spheroids bounded 
by the external pressure exerted by the parent cloud 
\citep[e.g.][]{Ward-Thompson+94,Alves+01,Tafalla+04,JKirk+05,Roy+14}. 
These BE-like density profiles do not imply that prestellar cores are necessarily in hydrostatic equilibrium 
and are also consistent with dynamical models \citep{Ballesteros+03}.

The {\it Herschel} space observatory \citep{Pilbratt+10} 
has led to a revolution in submillimeter dust continuum imaging 
and therefore surveys for dense cores, 
thanks to its unprecedented mapping speed, high sensitivity and dynamic range, multi-wavelength coverage (from $70\, \mu$m to  $500\, \mu$m), 
and reasonably high resolution (18\arcsec ~at $\lambda = 250\, \mu $m, corresponding to $\sim 0.03$~pc at $d = 350$ pc). 
This has allowed the whole extent of nearby molecular clouds to be searched for dense cores 
as part of the {\it Herschel} Gould Belt survey \citep[HGBS --][]{Andre+10} 
and clear connections to be made between 
the core formation process and the filamentary structure of the parent clouds (see \S ~3.2 below). 
In addition to providing deeper, more complete samples of dense cores, key advantages of {\it Herschel} surveys 
over earlier ground-based observations have been 1) the direct determination of core temperatures through multi-wavelength data, 
and 2) the ability to identify the boundaries of individual cores 
based on sensitive (radial) density profiles (cf. Fig.~\ref{core_profiles})  as opposed to arbitrary low signal-to-noise cutoffs.

With more sensitive, bigger datasets, 
the task of extracting candidate dense cores in wide-field submillimeter dust continuum images 
of highly structured molecular clouds has been both facilitated thanks to higher signal-to-noise data  
and made more complex owing to the larger amount of data to process and the broader range 
of cloud structures effectively probed. 
Multi-wavelength {\it Herschel} continuum data with resolution depending linearly on wavelength 
have required sophisticated, dedicated approaches such as \textsl{getsources} to identify candidate dense cores \citep{Menshchikov+12}. 
Briefly, the core extraction process with \textsl{getsources} consists of a detection and a measurement stage. 
At the detection stage, \textsl{getsources} analyzes fine spatial decompositions of the original maps at all observed 
wavelengths over a wide range of scales using a series of unsharp-masking steps. 
This decomposition makes it possible to filter out irrelevant spatial scales and to identify, for each object, 
the optimum scale at which it is visible in the data, greatly improving source detectability, 
especially in crowded regions and for extended sources. 
The multi-wavelength design of \textsl{getsources} also combines data over all wavelengths and 
produces a wavelength-independent detection catalog. 
At the measurement stage, 
the properties of each detected source are measured in the original (unfiltered) observed images. 
Two alternative methods that have also been used on {\it Herschel} data including \textsl{csar} \citep{JKirk+13}, 
a conservative variant of the well-known segmentation routine \textsl{clumpfind} \citep{Williams+94}, and 
\textsl{cutex} \citep{Molinari+11}, 
an algorithm that identifies compact sources by analyzing multi-directional 
second derivatives and performing ``curvature'' thresholding in monochromatic images. 

Once cores have been extracted from the maps, the {\it Herschel} observations provide a very sensitive way of distinguishing between 
protostellar cores and starless cores based on the presence or absence of point-like 70~$\mu$m emission. 
Indeed, the emission flux at 70~$\mu$m traces very well the internal luminosity of a protostar \citep[e.g.][]{Dunham+08}, 
and {\it Herschel} observations 
of nearby ($d < 500$~pc) clouds have the sensitivity to detect even candidate ``first hydrostatic cores''  \citep[cf.][]{Pezzuto+12}, 
the very first and lowest-luminosity ($\sim 0.01$--0.1$\, L_\odot $) stage of protostars 
\citep[e.g.,][]{Larson69, Saigo+11, Commercon+12}.

The {\it Herschel} continuum data can also be used to divide the sample of starless cores into gravitationally bound 
and unbound objects based on the locations of the cores in a mass verus size diagram \citep[cf.][]{Motte+01} 
and comparison of the derived core masses  
with local values of the Jeans or BE mass (see Fig.~7 of  \citealp{kony2015}). 
The prestellar cores identified with {\it Herschel} have typical average volume densities 
$\sim 10^4$--$10^6\, {\rm cm}^{-3}$ and outer sizes between $\sim 0.01\,$pc 
and $\sim 0.1\,$pc. It is noteworthy that the {\it measured} core sizes are found to be smaller than the typical sonic scale 
$\sim 0.1\,$pc in the low-density ($< 10^4\, {\rm cm}^{-3}$) parts of molecular clouds, 
as expected (see above).

\subsection{Link to the filamentary structure of molecular clouds} \label{sec:CMF_obs_filament}

Thanks to the high surface-brightness sensitivity and spatial dynamic range achievable from space, 
a big step forward with {\it Herschel} imaging surveys compared to earlier submillimeter ground-based observations 
has been the ability to {\it simultaneously} probe compact structures such as dense cores {\it and} larger-scale structures 
within the parent clouds such as filaments. This has provided, for the first time, an unbiased view of both the spatial 
distribution of dense cores 
and the link between dense cores and the texture of molecular clouds. 
In particular, {\it Herschel} GBS observations have shown that most ($75\%_{-\,5\%}^{+15\%}$) prestellar cores 
are located within filamentary structures of typical column densities $N_{H_2} \simgt 7 \times 10^{21}\, {\rm cm}^{-2}$,  
corresponding to visual extinctions $A_V \simgt 7 $ (e.g. \citealp{Andre+10,kony2015,Marsh+16}; 
see also Fig.~\ref{taurus_cores}). 
Moreover, most prestellar cores lie very close to the crest of their parent filament 
\citep[e.g.][]{Bresnahan+18,Konyves+2020,Ladjelate+20}, 
that is within the flat  inner $< 0.1\,$pc portion of the filament radial profile \citep[cf.][]{Arzoumanian+11,Arzoumanian+19}. 

The column density transition above which prestellar cores are found in filaments is quite pronounced. 
It resembles a smooth step function as illustrated in Fig.~\ref{cfe_cmf}a, which shows the observed core formation efficiency 
$ {\rm CFE_{obs}}(A_{\rm V}) = \Delta M_{\rm cores}(A_{\rm V})/\Delta M_{\rm cloud}(A_{\rm V}) $ 
as a function of the ``background'' column density of the parent filaments in the Aquila cloud complex \citep{kony2015}.
There is a natural interpretation of this sharp column density transition for prestellar core formation 
in terms of simple theoretical expectations for the gravitational instability of nearly isothermal gas cylinders. 
Adopting the typical inner width $W_{\rm fil} \sim 0.1$~pc measured  for nearby molecular filaments with {\it Herschel} \citep[][]{Arzoumanian+11,Arzoumanian+19} 
and using the relation $M_{\rm line} \approx \Sigma_0 \times W_{\rm fil}$ between the central 
gas surface density $\Sigma_0$ and the mass per unit length $M_{\rm line}$ of a filament, 
there is a very good match between 
the transition at $A_V^{\rm back} \sim 7$ or $\Sigma_{\rm gas}^{\rm back} \sim $~150~$M_\odot \, {\rm pc}^{-2} $ 
and the critical mass per unit length 
$M_{\rm line, crit} = 2\, c_{\rm s}^2/G \sim 16\, M_\odot \, {\rm pc}^{-1} $  
of isothermal long cylinders in hydrostatic equilibrium for a sound speed $c_{\rm s} \sim 0.2$~km/s, i.e., 
a typical gas temperature $T \sim 10$~K 
\citep[e.g.][]{Ostriker64}. 
Therefore, the observed column density transition essentially corresponds to thermally transcritical filaments 
with masses per unit length within a factor of 2 of $M_{\rm line, crit} $, which are 
prone to gravitational fragmentation along their length \citep{InutsukaMiyama1992, Inutsuka+97, Fischera+12}.

\setlength{\unitlength}{1cm}
\begin{figure}
\begin{picture} (0,4.75)
 \put(0,0){\includegraphics[width=11.5cm,angle=0]{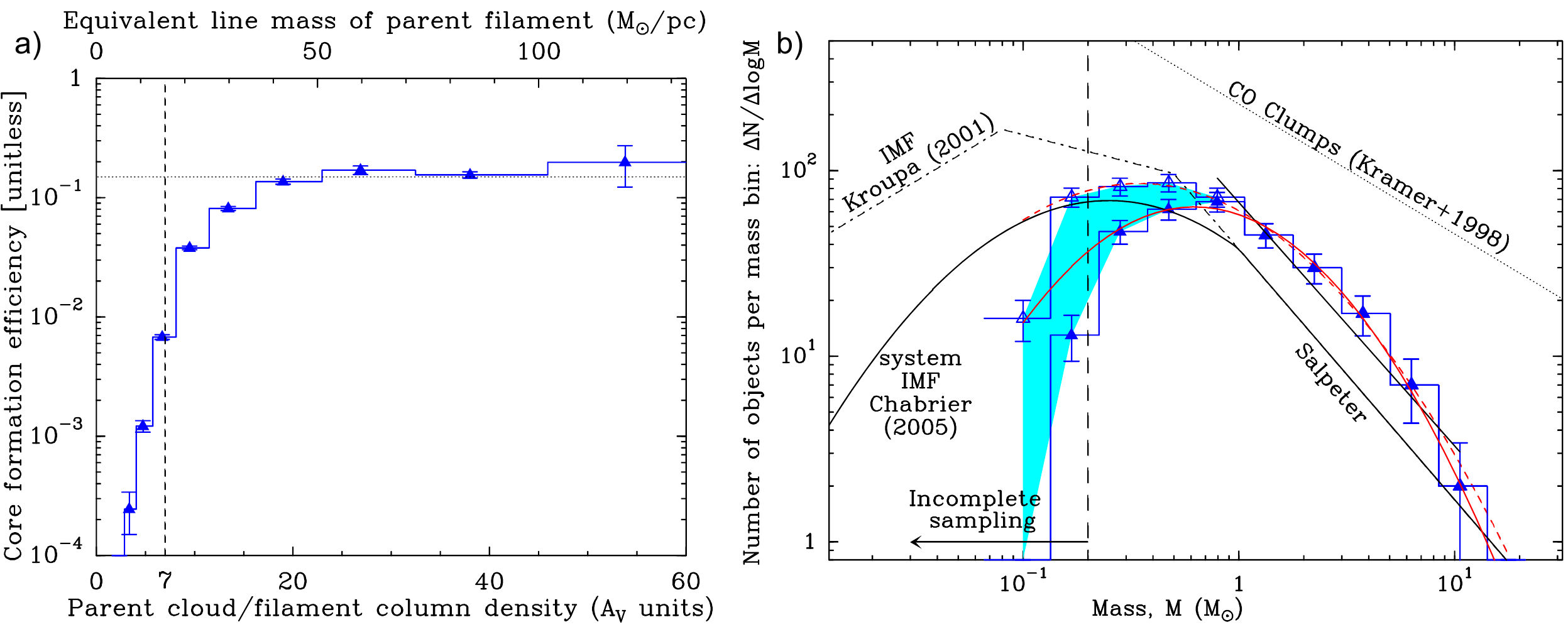} }
\end{picture}
 \caption{
a) Prestellar core formation efficiency 
[$ {\rm CFE}(A_{\rm V}) = \Delta M_{\rm cores}(A_{\rm V})/\Delta M_{\rm cloud}(A_{\rm V}) $]
in the Aquila molecular cloud as a function of background column density expressed in $A_{\rm V}$ units 
(blue histogram with error bars).  
b) Prestellar core mass function (in $\Delta N$/$\Delta$log$M$ format) derived from {\it Herschel} Gould Belt survey data 
in Aquila for a total of 446 candidate prestellar cores (blue histograms and triangles).
The blue shaded area reflects uncertainties in the selection of self-gravitating prestellar cores 
among the identified starless dense cores.
Lognormal fits (red curves) and a power-law fit to the high-mass end of the CMF (black solid line) are superimposed.
(Both panels adapted from \citealp{kony2015}.)
 }
\label{cfe_cmf}
\end{figure}

The observed spacing of dense cores along filaments is not consistent with the predictions of 
standard cylinder fragmentation theory, however. 
Linear fragmentation models for infinitely long, isothermal equilibrium  cylinders 
predict a characteristic core spacing of $\sim \,$ $4\, \times\, $ the filament width 
\citep[e.g.,][]{InutsukaMiyama1992}. 
In contrast, the spacing observed between {\it Herschel} prestellar cores is generally not periodic 
and the median value of the projected core separation is found to be close to the typical $\sim 0.1\,$pc 
inner width of filaments \citep[e.g.][]{Andre+14,Konyves+2020}. 
A few good examples of quasi-periodic chains of dense cores have also been found (see, e.g., Fig.~\ref{taurus_cores} and 
\citealp{Tafalla+15}), 
but again the corresponding characteristic spacing appears to be comparable to, rather than $\sim 4\, \times\, $ larger than, 
the diameter of the parent filament.
Moreover, complementary high-resolution studies with, e.g., interferometers 
\citep[][]{Takahashi+13,Teixeira+16,Kainulainen+13,Kainulainen+17,Shimajiri+19} 
have provided evidence of two distinct fragmentation modes within at least some thermally supercritical filaments: 
a) a ``cylindrical'' fragmentation mode corresponding to clumps or groups of cores 
with a separation consistent with $\sim 4\, \times\,$ the filament width given projection effects, 
and b) a ``spherical'', Jeans-like fragmentation mode corresponding to a typical spacing $\simlt 0.1\,$pc 
between cores (and within groups). 
This discrepancy between observations and simple theoretical predictions may be understood
by realizing that real molecular filaments are not isolated cloud structures in perfect hydrostatic equilibrium. 
In particular, the process of filament fragmentation is modified 
by the presence of accretion and low levels of turbulence \citep[][]{Clarke+16,Clarke+17}. 
Interestingly, \citet{Clarke+16} 
showed that the fastest growing mode 
for density perturbations in an accreting filament moves to shorter wavelengths, 
and can thus become significantly shorter than $\sim 4\, \times\, $ the filament width, 
when the accretion rate that the filament experienced during its formation  
is comparable to that observed in some filaments 
such as the Taurus B211/B213 filament ($\sim \,$27--50$\, M_\odot \,$pc$^{-1}\,$Myr$^{-1}$ -- \citealp{Palmeirim+13}).
While further theoretical work including, e.g., the role of magnetic fields, is still needed to get a perfect match with observations, 
this is very promising.

Overall, the observational findings summarized here  support a filament paradigm for low-mass star formation,  
in two main steps \citep{Andre+14,Inutsuka+15}:
first, $\sim \,$ 0.1-pc-wide molecular filaments are assembled, presumably 
through large-scale compression of cold interstellar material in supersonic MHD flows; 
second, gravity takes over and fragments the densest filaments with $M_{\rm line} $ near or above $M_{\rm line, crit} $ 
into prestellar cores and then protostars.

\subsection{Observations of the Core Mass Function (CMF)}\label{sec:CMF_obs}

The dense cores identified in  
submillimeter continuum imaging surveys of nearby molecular clouds
may represent the local mass reservoirs out of which individual stars form. 
Indeed, a sizable fraction ($\simgt 10\%$) of them 
harbor a central young stellar object
and are known to be protostellar in nature, while observed prestellar cores 
are morphologically very similar to protostellar cores (cf. Fig.~\ref{taurus_cores}b), suggesting a genetic link between 
the former and the latter. 
Moreover, as first pointed out by \citet{Motte+98} 
in the case of the Ophiuchus main cloud (L1688) more 
than two decades ago, the mass distribution of prestellar cores or prestellar {\it core mass function} (CMF) 
broadly resembles the stellar IMF in shape. 
Similar results were subsequently reported by a number of independent groups 
\citep[e.g.,][]{Testi+98,Johnstone+00,Johnstone+01,Motte+01,Stanke+06,Nutter+07,Enoch+08} 
in nearby star-forming regions such as Ophiuchus, Serpens, Orion A \& B, and Perseus. 
In all of these clouds, the observed prestellar CMF is consistent with the \citet{Salpeter55} 
power-law IMF at the high-mass end ($\Delta N/\Delta \log M \propto M^{-1.35}$), 
and significantly steeper than the mass distribution of both molecular clouds and diffuse CO clumps 
($\Delta N/\Delta \log M \propto M^{-0.7}$  -- e.g. \citealp{Blitz93, Kramer+1998}).

Pre-{\it Herschel} submillimeter continuum findings on the prestellar CMF were nevertheless limited by 
small-number statistics (with typically $< 100$ cores in any given study) and relatively large uncertainties in core masses 
due to, e.g., rather arbitrary assumptions about the dust temperature in the cores. 
The {\it Herschel} Gould Belt survey (HGBS -- \citealp{Andre+10}) 
was specifically designed to improve on this situation,   
by providing an essentially complete census of dense cores with well characterized temperatures, masses, 
and density profiles in most, if not all, nearby molecular clouds. 
The prestellar CMFs derived from HGBS data are generally lognormal in shape 
and resemble the \citet{Chabrier05} 
system IMF. In the Aquila cloud, for instance, the CMF found by \citet{kony2015} 
based on a sample of $\sim \,$450 candidate 
prestellar cores, complete down to a $> 90\% $ completeness level of $\sim 0.2\, M_\odot $, 
is well fit by a lognormal distribution which peaks at $\sim \,$0.4--0.6$\, M_\odot $ 
and has a standard deviation of $\sim 0.52\pm0.05 $ in log$_{10}M$, 
compared to a peak at $0.25\, M_\odot $ and a standard deviation of 0.55 
for the \citet{Chabrier05} 
system IMF (cf. Fig.~\ref{cfe_cmf}b). 
The Aquila CMF is consistent with an essentially one-to-one mapping between prestellar core mass 
and stellar system mass ($M_{\star \rm sys} = \epsilon_{\rm core}\, M_{\rm core} $) 
with a core-to-star formation efficiency $ \epsilon_{\rm core} \sim  0.4^{+0.2}_{-0.1}$ 
(see also \citealp{Alves+07}). 
Very similar results (with similar sample sizes and mass completeness levels $\sim 0.2$--$0.4\, M_\odot $) 
have recently been obtained for the prestellar CMFs of Cepheus and Perseus, 
by Di Francesco et al. (submitted) and Pezzuto et al. (submitted), respectively, also based on {\it Herschel} GBS data. 
In the nearest molecular clouds of the Gould Belt  at $d \sim 140\,$pc, such as 
Taurus, Corona Australis, Lupus, Ophiuchus, the HGBS census of prestellar dense cores has reached  
a deeper mass completeness level of $\sim 0.1\, M_\odot $, but the core number statistics are lower 
as these clouds are significantly less massive and contain much less dense gas. 
The prestellar CMFs found in these regions are nevertheless broadly consistent with those 
derived in Aquila, Cepheus, Perseus, given rather large statistical uncertainties, 
with tentative evidence of a somewhat lower peak mass at $\sim  \,$0.3--0.4$\, M_\odot $ 
suggestive of a somewhat higher  efficiency $ \epsilon_{\rm core} \sim  0.5$--0.7 
\citep{Marsh+16,Bresnahan+18,Benedettini+18,Ladjelate+20}.
{\it Herschel} data in combination with slightly higher angular resolution (12\arcsec) ground-based data 
taken at 1.1~mm with the Large Millimetre Telescope have also allowed the CMF to be derived 
in somewhat more distant and massive cluster-forming hubs such as Mon R2 ($d \sim 830\,$pc), with 
results consistent with those obtained in Gould Belt clouds \citep[][]{Sokol+19}.

Interestingly, the most massive prestellar cores identified with {\it Herschel} 
(with masses between $M \sim 2\, M_\odot$ and $\sim 10\, M_\odot$)
tend to be spatially segregated in the highest column density parts/filaments 
of the clouds, suggesting that the prestellar CMF is not homogeneous {\it within} a given cloud
but depends on the local column density (or line mass) of the parent filament 
(\citealp{Konyves+2020}; see also \citealp{Shimajiri+19}). 
In Orion~B, for instance, there is a marked trend for the prestellar CMF 
to broaden and shift to higher masses in higher density areas \citep{Konyves+2020}. 
A related point is the finding of ``top-heavy'' CMFs (with $dN/d \log M ~\sim M^{-1}$) 
in very dense clumps and filaments such as W43-MM1 imaged with ALMA in massive 
star-forming complexes beyond the Gould Belt \citep{Motte+18}. 
These recent results support the view that the global prestellar CMF results 
from the superposition of the CMFs produced by individual filaments \citep{lhc, Andre+19}.

In addition to prestellar cores, a large number of gravitationally unbound starless cores are detected 
with {\it Herschel}, especially in the nearest clouds. The mass function of these unbound  starless cores 
extends well below the peak of the prestellar CMF, approximately 
as a power law approaching that of the CO clump mass spectrum 
($\Delta N/ \Delta \log M \propto M^{-0.6}$, down to $M < 0.01\, M_\odot $ in Taurus --  \citealp{Marsh+16}). 

The {\it Herschel} results have therefore confirmed the existence of a close similarity between the 
prestellar CMF and the stellar IMF, with an order-of-magnitude better core statistics and more accurate 
core masses thanks to direct dust temperature estimates than earlier ground-based studies. 
The typical efficiency factor $ \epsilon_{\rm core} \sim 40\% $ or shift between the CMF and the IMF 
may be attributed to mass loss due to the effects of outflows during the protostellar phase \citep{Matzner+00,Machida+12,offner2017}. 
Some uncertainties remain 
concerning the exact location of the CMF peak 
and its detailed dependence on environment, as well as the low-mass end of the prestellar CMF 
below the peak, including the brown dwarf mass regime. 
The mere existence of a peak in the prestellar CMF has been questioned on the grounds 
that it lies only a factor of $\sim \,$2--3 above the mass completeness limit of current observational surveys (cf. Fig.~\ref{cfe_cmf}b)   
and that it tends to shift to lower masses when resolution is increased in numerical simulations \citep[e.g.][]{Ntormousi+19}. 
The latter does not seem to be true in observations, however, since high-resolution interferometric studies 
of nearby regions with, e.g., ALMA or NOEMA have failed to find a large number of pre-brown dwarfs, i.e., ultra-low-mass prestellar cores 
in the brown-dwarf mass regime. Some good examples of candidate pre-brown dwarfs exist \citep[e.g.][]{Andre+12}, 
but they appear to be quite rare compared to solar-type cores, which supports the presence of a peak in the prestellar CMF. 
Moreover, when observed at higher resolution with interferometers, {\it Herschel} dense cores typically show very little substructure 
and do not split up into many smaller-scale condensations but remain single or at most double objects 
\citep{Schnee+10, Maury+10,Dunham+16, Sadavoy+17,Kirk+2017,Maury+19}.
While more work will be needed to fully assess 
uncertainties in core mass estimates and 
the potential importance of subtle observational biases such as background-dependent incompleteness effects 
and blending of unresolved groups of cores, 
the overall shape of the prestellar CMF now seems reasonably well established by observations, 
at least for core masses $\sim \,$0.1--10$\, M_\odot $.

The simple idea of a direct {\it physical} connection between the prestellar CMF and the stellar IMF 
is thus tempting but remains debated 
\citep[e.g.][]{Ballesteros+06,Clark+07,Offner+14}.
In particular, it should be kept in mind that observed CMFs represent {\it snapshots} 
and that prestellar cores may evolve and, e.g., grow in mass before collapsing to protostars. 
In fact, the many unbound starless cores detected with {\it Herschel} provide evidence 
of a significant core building phase (cf. Fig.~8 of \citealp{Andre+14}), 
reminiscent of the process seen in 
some numerical simulations \citep[e.g.,][]{Gong+09,gong2015}. 
Despite such caveats, the {\it Herschel} findings summarized above tend to support models of the IMF based on pre-collapse cloud fragmentation 
such as the gravo-turbulent fragmentation picture 
\citep[e.g.][]{Larson85,Klessen+00,padoan1997,padoan2002,hc08}. 
More precisely, given that prestellar cores are predominantly found in dense molecular filaments (see \S ~\ref{sec:CMF_obs_filament} above), 
the {\it Herschel} observations suggest that the filamentary structure of molecular clouds plays an important role 
in shaping the prestellar CMF and by extension the stellar IMF.  
Indeed, \citet{Andre+14,Andre+19} 
proposed that the peak of the prestellar CMF results from gravitational fragmentation 
of transcritical, 0.1-pc-wide filaments and that the high-mass end of the CMF may be directly inherited from 
the observed Salpeter-like distribution of supercritical filament masses per unit length.

\section{Theory of the Core Mass Function (CMF)}
As recalled above, the shape of the CMF, which has been inferred from  observations
\citep[e.g.,][]{kony2015} appears to be similar to the shape of the IMF with nevertheless 
an important difference regarding the position of the peak. While the 
peak of the system IMF is about 0.2 $M_\odot$ \citep[e.g.,][]{Chabrier03}, it is about 2-3 higher 
for the CMF. This difference is attributed to the inefficiency of the core to 
star conversion. 
Thus a number of works have examined the origin of the CMF with the hope that it would constitute a valuable explanation for the origin of the IMF. While reasonable, this latter assumption remains controversial in particular because it assumes that  {\it cores}, defined usually with simple gravitational boundedness in theories, do not fragment or do it in a self-similar way. 

Here we first describe the various analytical approaches that have been developed 
to infer the CMF. We then describe the numerical simulations performed 
to study the CMF and the conclusions that have been drawn. 

\subsection{Analytical approach of the CMF}

\subsubsection{Simple considerations}

Before presenting the various theories that have been developed to explain the origin 
of the CMF, it is worth starting with some very general and simple considerations.
 In 3D the number of density fluctuations of wavenumber $k$ is expected to be such that  $N(k)\propto k^3$ 
simply because $k^3$ is the volume in $k$ space.
However, the mass associated with the fluctuation of
scale $R$ is $M=\rho R^3\propto k^{-3}$, leading to 
\begin{eqnarray}
\frac{dN}{dM}\propto M^{-2}.
\end{eqnarray}
or equivalently
\begin{eqnarray}
\frac{dN}{d \log M}\propto M^{-1}.
\label{geo}
\end{eqnarray}
The mass spectrum (\ref{geo}) thus corresponds to the most natural mass distribution, set up by fluctuations 
obeying purely geometrical considerations. 
While obviously extremely elementary, this has the advantage of showing why 
it is not surprising that mass spectra in the interstellar medium exhibit a power-law behaviour with an 
exponent close to 2, namely $d N / dM \sim M^{-1.6}$ for unbound structures such as CO 
clumps, $dN / dM \sim M^{-2.0} $ for stellar clusters, and $dN / dM \sim M^{-2.35}$ for cores/stars. 
In this respect, it is not so surprising that many theories of the IMF have derived mass spectra that present power-spectrum behaviour with an exponent close to -2.
It is essential when developing a theory that explains the CMF or the IMF to 
predict the exponent with enough accuracy and to have full consistency of the 
physical assumptions.  

\subsubsection{The CMF from supersonic turbulence}\label{sec:CMF_theory}

The first theory combining supersonic turbulence and self-gravity was developed
 by \citet{padoan1997}. The authors assumed
a lognormal density distribution  as  inferred from
 numerical simulations \citep[e.g.][]{vazquez1994,kritsuk2007,fedban2015}.
 The regions of the flow that 
are thermal Jeans unstable are then selected as the core masses. 
This approach led them to infer a mass spectrum that presents as a 
power law at high masses
 as a result of the scaling relation between local density and Jeans mass
(typically $dN/ d \log M \propto M^{-2}$).
Interestingly, at low masses, they infer a lognormal shape, a direct consequence of the lognormal density distribution.

The model proposed later by \citet{padoan2002} is quite different. 
They envisioned  a shocked layer  in a weakly 
magnetized medium. They further required that the magnetic field be 
 perpendicular to the incoming velocity field. This led them 
to a CMF that has the correct shape. As discussed below, it is
clear, however, from various hydrodynamical simulations that the magnetic field does not modify the CMF or the IMF significantly (see \S~ \ref{sec:IMF_simu}).

\citet{hc08}  proposed a more general theory  
\citep[see also][]{hc2013}. The approach 
consists in counting the mass of the fluid regions
within which gravity dominates over  thermal, turbulent and 
magnetic support, that is to say  fluid elements that are gravitationally unstable. 
In this theory, turbulence plays a dual role, on one hand
 it enhances star formation by locally compressing the gas, i.e., making a broader
density PDF,  but on the 
other hand, it also quenches star formation because of the turbulent 
dispersion that resists gravity.

The \citet{hc08} theory constitutes an extension of the mathematical formalism developed by \citet{press1974}, which was first adapted to star-formation context by \citet{Inutsuka2001}.
The most noticeable  differences are (i) the characteristics of the density fluctuations and (ii) the 
criteria that unstable regions must satisfy. 
The density fluctuations, which are 
small and Gaussian  in the cosmological case,  are assumed to be lognormal in the 
star formation case. In cosmology  the selection criterion is a simple  density 
threshold while it is  based on the energy equipartition in the second case
\begin{eqnarray}
  \langle V_{\rm rms}^2\rangle    + 3\, (c_{\rm s}^{\rm eff})^2 < - E_{\rm pot} / M.
\label{viriel_ceff}
\end{eqnarray}
\citet{hc08} assume a \citet{larson1981} 
 (see also \citet{hf2012})
 velocity-size relationship for $V_{rms}$ (although see \S 7.4 for a discussion of the validity of Larson's relations).
The turbulent rms velocity follows  a power-law correlation with the size of the region, 
\begin{eqnarray}
\langle V_{\rm rms}^2\rangle =  V_0^2 \times \left( {R \over  1 {\rm pc}} \right) ^{2 \eta},
\label{larson}
\end{eqnarray}
with $V_0\simeq 1\, {\rm km\, s}^{-1}$ and $\eta \simeq 0.4$-0.5.

In more detail, the calculations entail the following steps. First, the density field is smoothed at a 
scale, $R$, using a window function. Second, the  mass enclosed in  areas that, at scale $R$,  
have a density  larger than the  specified density  criterion $\delta_R^c$,  is inferred  from the density PDF. 
Due to mass conservation, this unstable mass at scale $R$ must be  equal to the  integrated mass of the  structures
whose mass is larger than a scale dependent critical mass $M_R^c$. Mathematically, this leads to the following 
expression
\begin{eqnarray}
 \int ^{\infty} _ {\delta_R^c} \bar{\rho} \exp(\delta)   {\mathcal P}_R(\delta)  d\delta
\label{hc_eq1} = \int _0 ^ {M_R^c} M' \, {\mathcal N} (M')\,   P(R,M')\, dM', 
\end{eqnarray}
where $\delta_R^c = log(\rho_R^c/\bar{\rho})$ 
and $\rho_R^c= M_R^c / (C_m R^3)$, with $C_m$ being a dimensionless coefficient of 
order unity, ${\mathcal P}_R$ is the density PDF, generally assumed to be lognormal, while 
$P(R,M')$ is the probability of having an  unstable mass, $M'$
embedded in $M_R^c$ at scale $R$. So far it has been assumed that $P(R,M')=1$.

Since Eq.~(\ref{hc_eq1}) is expressed as a function of the scale, $R$, one can take 
 its derivative  with respect to $R$ to  get  the mass spectrum
\begin{eqnarray}
\label{n_general}
 {\mathcal N} (M_R^c)  &=& 
 -{ \bar{\rho} \over M_R^c} 
{dR \over dM_R^c} \,
\ {d \delta_R ^c \over dR} \exp(\delta_R^c) {\mathcal P}_R( \delta_R^c).
\end{eqnarray}
Note that there should be  a second term that is important to explain the mass spectrum of unbound clumps defined 
by a uniform density threshold (such as observed CO clumps), but since it plays a minor role for 
(virial defined) bound cores, we drop it here.
Equation~(\ref{n_general}) nicely shows that the mass spectrum depends on i) the density PDF and \textcolor{Brown}{ii)} the 
mass-size relation,
which follows from the virial theorem. We also stress that so far no assumption 
has been made regarding the nature of the density PDF, which therefore does not need to be a lognormal.

If we assume a lognormal PDF, whose variance is given by $\sigma^2 = \ln (1 + b^2 {\mathcal M}^2)$,
then the  theory is controlled by two Mach numbers. 
First, we have
\begin{eqnarray}
{\mathcal M}_* = { 1  \over \sqrt{3} } { V_0  \over c_{\rm s}}\left({\lambda_J^0 \over   1 {\rm pc} }\right) ^{ \eta}
\approx (0.8-1.0) \,\left({\lambda_J^0\over 0.1\,{\rm pc}}\right)^{\eta}\,\left({c_{\rm s}\over 0.2\, {\rm km \, s^{-1} }}\right)^{-1},
\label{mach_eff}
\end{eqnarray}
defined as the  velocity dispersion to sound speed ratio at the mean Jeans length, $\lambda_J^0$ (and not at the local 
Jeans length). Note that ${\mathcal M}_*$ simply
represents the first term of Eq.~(\ref{viriel_ceff}).
Second, we define the  Mach number, ${\mathcal M}$,
 at the injection scale, $L_i$, which we assume to be the typical
 size of the cloud
${\mathcal M}={\langle V^2 \rangle^{1/2} / c_{\rm s}}$.

While the  Mach number, ${\mathcal M}$,  broadens the density PDF, 
which  tends  to promote star formation by 
creating new overdense collapsing gas, ${\mathcal M}_*$, which 
we recall plays a role through  energy equipartition, is the  additional non-thermal support
induced by the turbulent dispersion.
 At large scales  turbulence stabilizes the
 parcels of fluid that would be gravitationally unstable if thermal 
pressure were present as clearly shown by Eq.~(\ref{viriel_ceff}).

For ${\mathcal M}_* \ll 1$, that is to say when the turbulent dispersion is small 
with respect  to the thermal support, 
it can be inferred 
from Eq.~(\ref{n_general})  that
the CMF at large masses  is identical to the  result of \citet{padoan1997}
i.e., $dN/d\log\, M \propto M^{-2}$ (see \citet{hc08} for further details). When ${\mathcal M}_* \simeq 1$,
 $dN/d\log\, M \propto M^{-(n+1)/(2n-4)}$, where the index of the velocity powerspectrum  $n$
is related to $\eta$ by the relation $\eta=(n-3)/2$. Since numerical simulations have demonstrated 
that  $n \simeq 3.8-3.9$ \citep[e.g.,][]{kritsuk2007}, the predicted slope is very close to  Salpeter's 
estimate and equal to about 1.25-1.4.

\citet{hopkins2012} and \citet{hopkins2013} proposed a complementary formulation  using 
excursion set theory \citep{bond1991}, which is based on random walks  
 in the Fourier space of the density field.  
  Clouds are defined by the scale at which  
  density fluctuations cross a specific  {\it barrier}, that 
  is to say, they reach the scale-dependent density threshold relevant 
  for the problem under consideration. To study gravitationally 
  bound clouds, \citet{hopkins2013} also use the 
  Virial theorem as stated by Eq.~(\ref{viriel_ceff}).
They propose a global model that
includes spatial scales larger than that of molecular clouds, which describes the whole galactic disc.
An appealing concept is that the clouds defined as density fluctuations that first cross
 the barrier (that is to say when diminishing the spatial scale 
this is the first time, that self-gravity becomes dominant) 
present a mass spectrum
that is slightly shallower than $dN / d \log \propto M^{-1}$. 
On the other-hand the density 
fluctuations that cross the barrier for the last time 
(i.e., at small spatial scales, self-gravity never dominates again)
have a mass spectrum almost 
identical to the one inferred in \citet{hc08}.
The physical interpretation is as follows.
The first type of structure, that is to say the fluctuations that 
cross the barrier for the first time,  likely
 represent massive molecular clouds. They are not themselves
 embedded in a larger self-gravitating cloud, while the second type likely represents prestellar core progenitors. Indeed
 it suggests that, while large-scale self-gravitating clumps should
 have a mass spectrum close to $ d N / d \log M \propto M^{-0.8-1}$,  the mass spectrum of the
 smallest self-gravitating fluctuations, likely analogs of dense cores, should be close to the observed CMF.

\subsubsection{The CMF from filaments}\label{sec:CMF_theo_filament}
As suggested by recent observations \citep[see e.g.][]{Andre+14}, most dense cores, if not all,  appear to be located inside filaments. Thus, various efforts have been undertaken to understand core formation inside filaments.

The first approach to infer the CMF  from a 
filamentary cloud was proposed by \citet{Inutsuka2001} who considered 
a critical filament collapsing along its major axis but not radially.
In particular, the line-mass function  obtained by  \citet{Inutsuka2001} 
predicts $dN/dM \propto M^{-2.5}$  provided the density fluctuations follow $\delta^2 \propto k^{-1.5}$.
 \citet{Roy+2015} have recently obtained  the
 power spectrum of density fluctuations along  sub-critical filaments
 in the Gould Belt Survey. 
They infer that $\delta^2 \propto k^{-1.6}$, which is compatible with the value assumed by \citet{Inutsuka2001}.   
Note that strickly speaking the mass function discussed by \citet{Inutsuka2001} corresponds to the mass function of stellar systems, 
i.e., groups of stars, which may include binary or multiple stars, which are typically separated by 4-8 times the filament width. Technically, \citet{Inutsuka2001}
considered  the integral over the mass that appears in eq.~(\ref{hc_eq1}) from 
$M_R$ to $\infty$, which  corresponds to the first crossing  \citep{hopkins2012} and not from 
0 to $M_R$, which is the last crossing and should represent the smallest self-gravitating objects.

\begin{figure}
\centering
\includegraphics[width=9cm,angle=0]{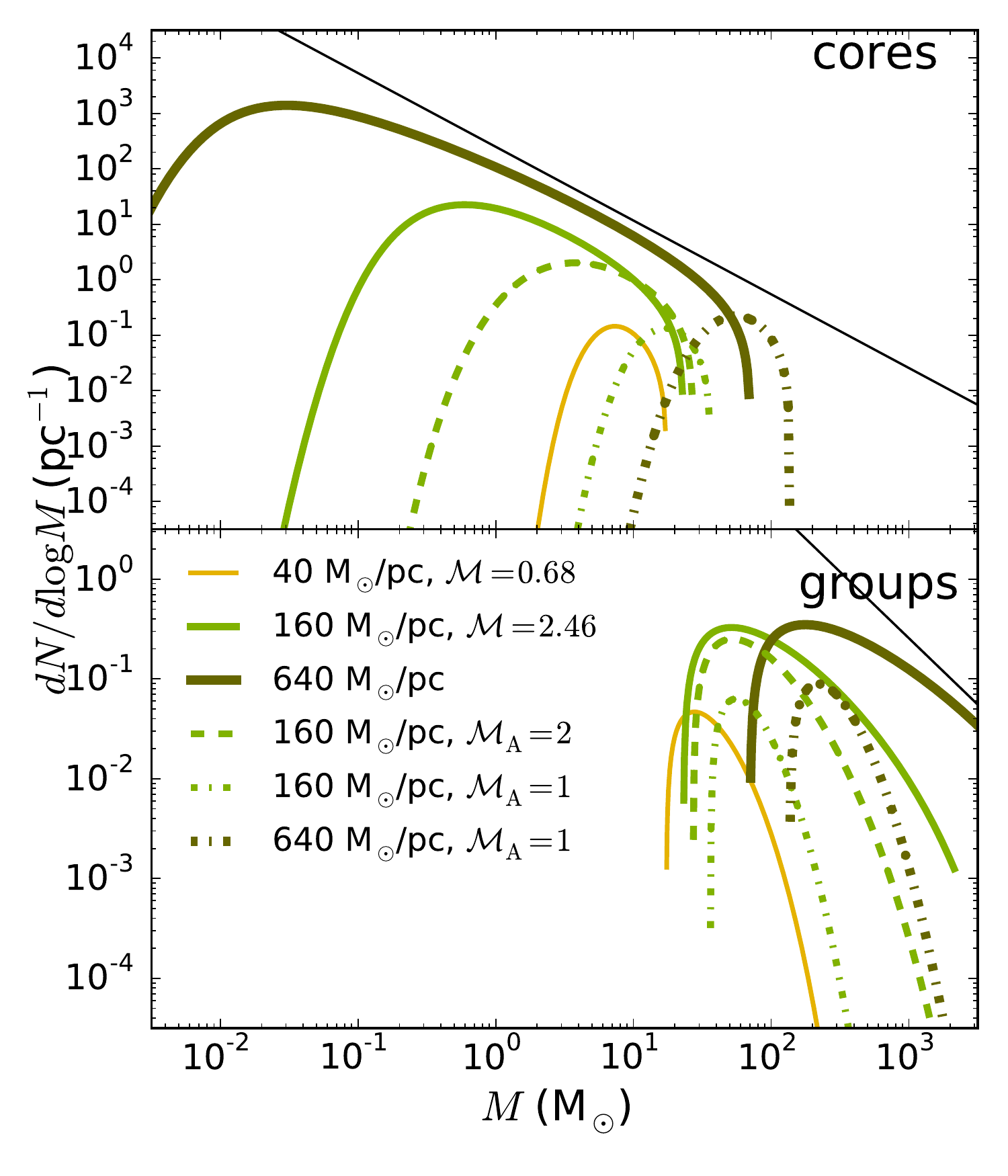} 
\caption{CMF inside filaments of different mass per unit length and magnetization. Solid curves show the CMF in non-magnetized filaments of 40 (yellow), 160 (light green), and 640 (dark green) $M_\odot$ pc$^{-1}$ with increasingly wide curves. Dashed and dot-dashed light green curves show the CMF in 160 $M_\odot$ pc$^{-1}$ filament with an Alfv\'enic Mach number,
$\mathcal{M}_ {\rm alfv} = 2$ and $1$. 
The dot-dashed dark green curve shows the CMF in a 640 $M_\odot$ pc$^{-1}$ filament with  $\mathcal{M}_ {\rm alfv} = 1$. 
The straight line shows the slope -1.33 measured by \citet{kony2015} in Aquila. Canonical values of 10 pc length and 0.1 pc diameter are used for all filaments. Figure extracted from \citet{lhc}.}
\label{lhc}
\end{figure}

\begin{figure}
\centering
\includegraphics[width=0.49\textwidth]{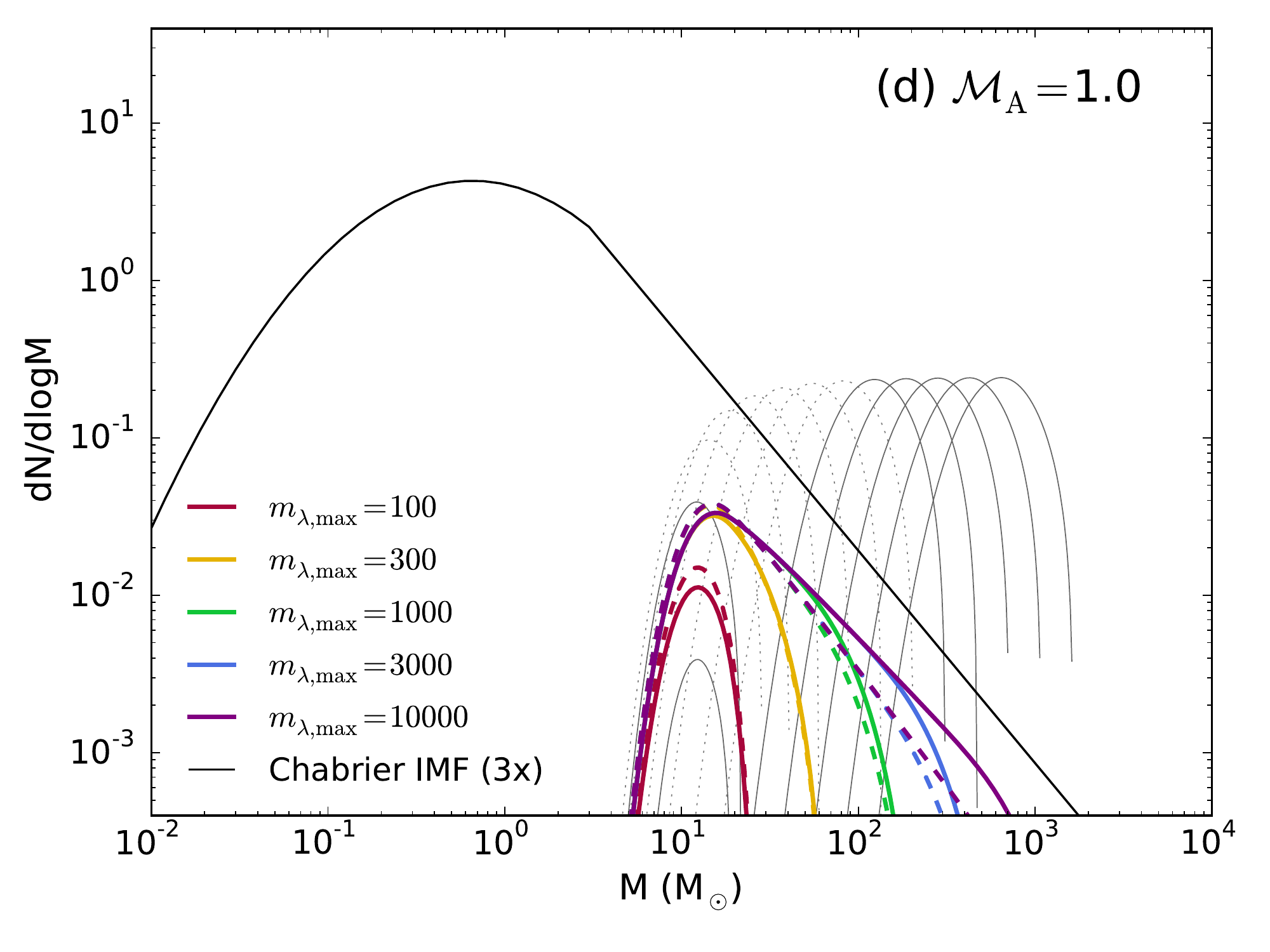}
\includegraphics[width=0.49\textwidth]{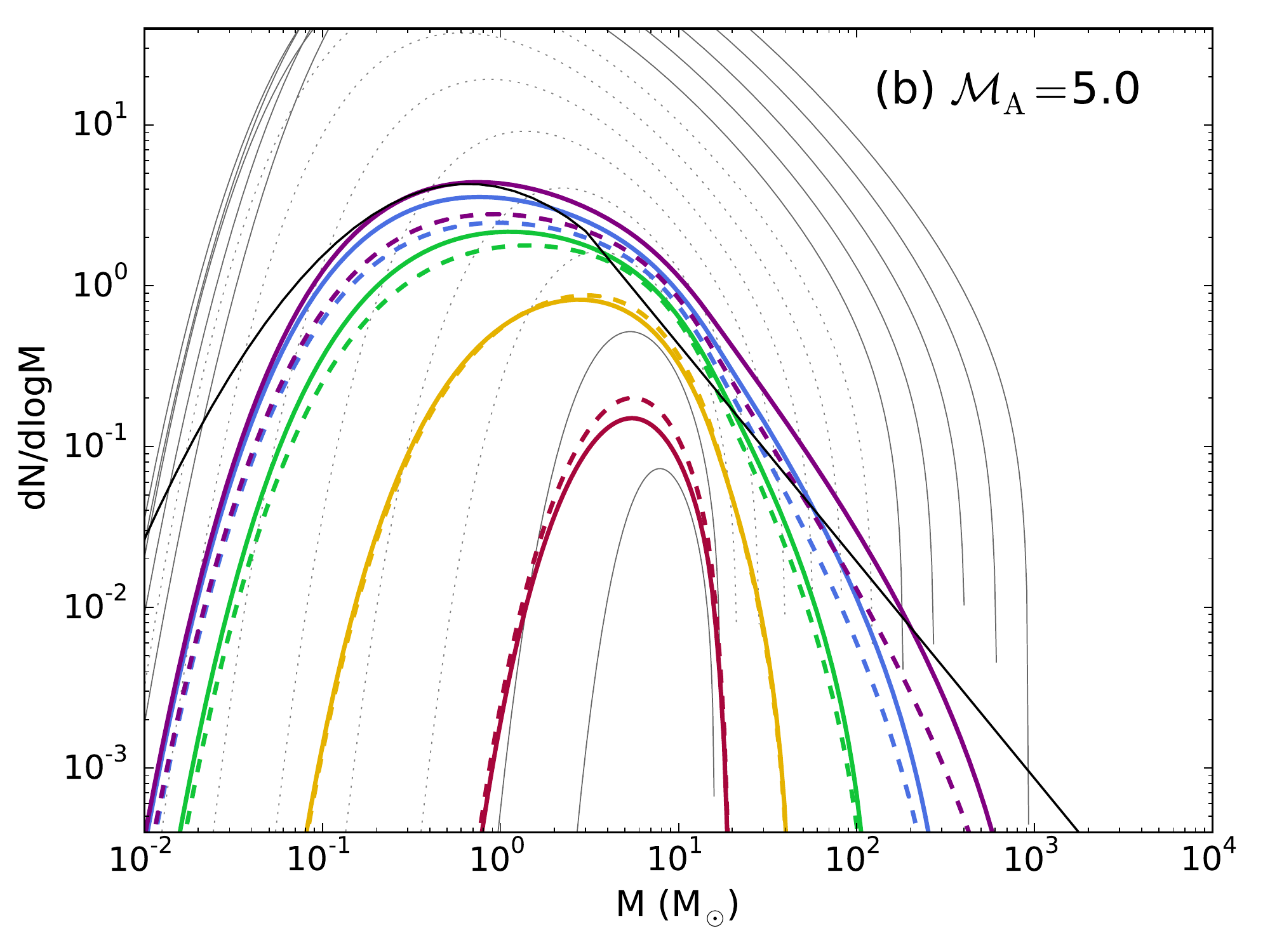}
 \caption{
Global CMF resulting from the sum of the individual CMFs of filaments having a mass per units length distribution $dN / d \log MpL \simeq MpL^{-1.2} {\rm (solid)~ 
or~}  MpL^{-1.5}{\rm ~(dashed)}$. The colors correspond to the summed CMF with different upper bounds of the mass per unit length as shown in the plot. Gray curves represent the CMF from individual filaments. 
A CMF in good agreement with observations is obtained for an Alfv\'enic Mach number of 5. Figure extracted from \citet{lhc}.}
 \label{lhc2}
\end{figure}

More recently, \citet{lhc} presented  an analytical theory for both the CMF and the mass function 
of groups of cores within  supercritical filaments, that is to say they considered the mass function obtained by estimating the mass integral  in eq.~(\ref{hc_eq1}) both between 0 and $M_R$ and between $M_R$ and $\infty$. 
The theory, which generalizes the calculations performed by \citet{Inutsuka2001} and \citet{hc08}, 
considers magnetized filaments that are radially supported by turbulent motions and have a constant width assumed to be close to 0.1 pc \citep{Arzoumanian+11,Arzoumanian+19}.
The model of \citet{Inutsuka2001} is a one-dimensional that considers fragmentation along the filament axis, while \citet{lhc} describe the one-dimensional fragmentation into groups and the 3-dimensional fragmentation into cores, that are smaller than the filament width. 
This picture is supported by observation of prestellar core distribution within filaments which shows regular grouping along the filament \citep[e.g.][]{hacar2013,Kainulainen+13,Takahashi+13}.
By assuming a functional dependence between the length and the radius of the density fluctuations, this approach describes the transition from small nearly spherical fluctuations to large and elongated ones.
While the former are not affected by the filamentary geometry of the cloud, the latter, on the contrary, are almost one dimensional objects and they only see the one-dimensional fluctuations along the filament axis.  

The  predicted CMF (see Fig.\ref{lhc}) is found to depend on the mass per unit lengths ($MpL$) and the magnetic intensity. 
In particular, it is found that in the absence of magnetic field, filaments with  high $MpL$ fragment into too many small cores. In the presence of a moderate magnetic field and for sufficiently high $MpL$, the CMF is compatible with the observed ones, that is to say it presents a peak around 0.5-1 $M_\odot$ in general. 
This however requires the Alfv\'enic Mach number, $\mathcal{M}_ {\rm alfv}$, to be on the order of 2-3. For lower values, typically 
$\mathcal{M}_ {\rm alfv} \simeq 1$, no low-mass core would develop and the peak of the CMF would be at values of 10 M$_\odot$ or more. 

Figure~\ref{lhc} shows that for filaments with low $MpL$, typically the nearly critical ones, 
the CMF is sharply peaked around the mean filament Jeans mass.
Both low- and high-mass cores are therefore absent. Consequently, the question arises how exactly the CMF is  determined
in the filament scenario. Is it established separately in most filaments or does it come from the contribution of a population of filaments with different $MpL$? Indeed \citet{Andre+14,Andre+19} found that 
filaments present a distribution of mass per unit length (filament line mass function, FLMF) that is proportional to $dN / d \log MpL \simeq MpL^{-1.5}$ in the $MpL$ range between $\sim 10$ and $\sim 200~M_\odot$/pc.

\citet{lhc} calculated the total CMF resulting from the individual CMF of 
 filaments having $MpL$ that follows the distribution  $dN / d \log  MpL \simeq MpL^{-1.5}$. Figure~\ref{lhc2} displays the results for two values of magnetic over turbulent energy corresponding to $\mathcal{M}_{\rm alfv} = ~5~ {\rm and~}1$. Stronger magnetic fields allow the formation of more massive fragments, while decreasing the total number of fragments and drastically reducing the number of low-mass ones. A striking result is that only when filaments with high $MpL$ are included does the convoluted CMF recover a Salpeter-like slope at large masses. 
This implies that magnetized massive filaments, with magnetic field energy in rough equipartition with turbulent energy, are necessary to reproduce the observed slope of the CMF. 
 Filaments with such high $MpL$ are statistically rare, while they are indeed observed in the massive DR21 ridge \citep[][$\sim 4000~M_\odot$/pc]{Hennemann+12} and in the W43-M11 cloud \citep[][$\sim 6000~M_\odot$/pc]{Motte+18}. 
When the magnetic field is too strong, the fragmentation is very limited and the peak of the CMF happens at values larger than the observed peak. 
For low levels of magnetization, the whole mass spectrum, including the low-mass end, is dominated by filaments of high $MpL$ that fragment excessively, 
and thus the peak position varies with the upper mass limit of integration.
At $\mathcal{M}_{\rm alfv} \simlt 5$, the impact of magnetic field becomes important enough for the peak mass to become basically independent of the integration over the filament population, producing reasonable agreement with \citet{Chabrier05} With an Alfv\'enic Mach number of a few, the model reproduces the scenario suggested by observations in \S~\ref{sec:CMF_obs}. 

\subsubsection{Effect of gas accretion}

Using pure $n-$body numerical simulations, \citet{Kuznetsova+17} have shown that groups of particles (clusters, in their picture) develop a mass function with slope of $-1$ and an $\dot{M} \propto M^2$ relation. This scaling relation comes from the gravitational focusing effect of the stellar potential that leads to competitive accretion. 
A similar idea was explored by \citet{Maschberger+2014, Bonnell+01}, while they found $\dot{M} \propto M^{2/3}$, as a result of gas-dominated gravitational focusing. Such relations can naturally lead to a powerlaw shape for the mass function. 
Since the simulations lack thermal physics, these authors concluded that gravity is responsible for developing a Salpeter-like slope and that some type of Bondi-Hoyle-Littleton accretion\footnote{see \S\ref{sec:BHL}} must occur due to gravity not only for stars, but also at cluster scales \citet{BallesterosParedes+15, Kuznetsova+17, Kuznetsova+18}. 

Motivated by these results \citet{VazquezSemadeni+19} made the conjecture that the connection between the CMF and the IMF is a natural consequence of a global, hierarchical and chaotic collapse of the molecular cloud \citep[see also][]{Zinnecker1984}: at all scales, objects accrete from their parent structures mainly due to gravity in a self-similar way. 
This simple result explains why numerical simulations with gravity tend to produce CMF and/or IMFs consistent with Salpeter slopes. On the contrary, when turbulent motions are strong, the density field fragments in excess, overpopulating the small scales \citep{KimRyu05}, i.e., producing an excess of small-scale cores and a deficit of large-scale ones \citep{BallesterosParedes+06}. In such cases, the resulting IMF shows only a few large-mass sinks and no low-mass ones \citep{Clark+08, BertelliMotta+16}. This is because only the few large cores that eventually form in such violent environments meet the conditions to also form a sink, namely, to be massive enough for gravity to overcome turbulent support.

On the other hand, some theories considered the accretion being stopped due to the exhaustion of the mass reservoir or the outflow from the protostar, therefore setting the final mass of the star.  These models either consider outflow counterbalancing the infall rate \citep{Adams1996}, a functional decay of mass accretion rate \citep{BasuJones2004, Myers2009}, or a probability for the core to be ejected from the initial mass reservoir \citep{Bate05a,Essex+2020}. These models are mostly based on adjustable statistical parameters. They have the advantage of being flexible to reproduce the observed CMF, while lack a precise link to the underlying physical mechanisms.

\subsection{Difficulties of the gravo-turbulent theories}
One natural question about any IMF theory is to which extent it varies with 
physical conditions. Indeed, there is strong observational support for a nearly invariant 
form and peak location of the IMF in various environments under Milky Way like conditions \citep[see e.g.][]{Offner+14}.
 Theories invoking the Jeans length like the ones presented above may 
have difficulties explaining the apparent universality of the peak since
 it is linked to the Jeans mass, which varies with gas density. 
Indeed, Fig.\ref{lhc} clearly shows that when different choices are made regarding the magnetic 
field or the filament mass per unit length, the integrated CMF from a filamentary region can be drastically different and the peak position may change significantly. 

Various solutions to this problem
have been proposed. For instance, \citet{Elmegreen+08} and \citet{Bate09b} argued  
that the gas temperature may be an increasing function of  density, leading to a 
Jeans mass that weakly depends on the density, while \citet{h2012} and \citet{lee2016b} proposed that 
for clumps that satisfy Larson's relations, there is competition between the Mach number dependence  of the density PDF and the density dependence of the Jeans mass, eventually leading to a peak mass that is insensitive to the clump
size.

A related problem comes from the density PDF. In the theories discussed above, it has been assumed to be lognormal. However in a collapsing cloud, this is not the case. The density PDF develops a
high-density tail, which is typically $\propto \rho^{-1.5}$ \citep[e.g.,][]{kritsuk2011}. In this case, the CMF predicted by gravo-turbulent models does not present a peak, instead a scale-free mass spectrum is inferred from a scale-free density PDF  \citep{lee2018a}. Therefore, the origin of the  CMF peak (and therefore the IMF peak if one assumes that they are linked) remains unclear.

\subsection{Modeling the CMF in numerical simulations}\label{sec:CMF_simu}
Various numerical simulations have 
simulated the formation of dense cores in 
molecular clouds. A broad diversity of initial conditions, physical processes, numerical 
setup and techniques have been used, so we do not review them here. The cores themselves 
have to be identified in the simulations and usually this is done in 
post-processing using a clump finding algorithm. 
In a second phase, the amount of thermal, magnetic and possibly 
even turbulent support must be estimated, and only unstable dense cores are usually selected. 

The CMF has been inferred by \citet{BallesterosParedes+06}, \citet{tilley2007}, \citet{nakamura2008}, \citet{nakamura2011}, \citet{chen2014} and \citet{h2018}.
In most cases the inferred CMF 
is similar to the observed CMF \citep{kony2015} (although \citet{BallesterosParedes+06} argue that the shape may depend on the Mach number, since larger Mach numbers fragment the medium in a more vigorous way, modifying the density power spectrum of the cloud \citep{KimRyu05}). Typically,
the simulated CMF exhibits a peak and a powerlaw at large masses with an exponent that is usually compatible  with the observed one \citep{tilley2007,nakamura2011,h2018}. These CMFs are therefore consistent with the scenario in which core formation is due to gravity and turbulence support while magnetic fields do not have a 
significant impact \citep[e.g. Fig.11 of][]{nakamura2011}. 

The question of the peak  remains controversial. 
As discussed in \S\ref{sec:CMF_obs}, the observed peak of the CMF 
is typically around 0.5-1 $M_\odot$. In numerical simulations the peak of the CMF is problematic for several reasons. First, 
 isothermal simulations with ideal MHD have no preferred scale and 
can be rescaled at will. Thus the peak position is not well determined. It relies 
on the initial conditions, which in principle can vary significantly.
Second, the issue of numerical convergence must be thoroughly checked.
 In the simulations presented in  \citet{h2018}, it has been found that  
 the peak of the CMF varies with numerical resolution.  In contrast, \citet{gong2015}
 concluded that in their colliding flow calculations, the peak is robust and does not depend 
on numerical resolution. 
While the reason for this apparent contradiction remains to be clarified, 
it is likely due to 
differences of the physical conditions studied.
In highly bound regions,  gravity 
induces a density PDF with a high-density powerlaw tail in which case the real CMF may not present a 
peak at all (see discussion in \citet{lee2018a}). 
Therefore, the appearance and location of the CMF peak depends on the numerical resolution of the simulation. 
On the contrary, in clouds not dominated by gravity, the PDF tends to be lognormal in which case the 
CMF displays a real peak and therefore convergence is possible.

\section{The Initial Mass Function (IMF): theory and simulations}

\subsection{The characteristic mass of the IMF: a peak}\label{sec:IMF_peak}

Several theories have been proposed to explain the origin of the characteristic or peak mass of the IMF. Here, we describe four different models.
Since star formation often involves very hierarchical collapse that produces subsequently smaller and smaller fragments, many models for the peak mass concern ways to impede fragmentation and impose a lower limit on the star-formation mass. The propositions discussed below are not necessarily mutually exclusive. It is possible that the coincidence of multiple mechanisms actually leads to a very similar IMF peak mass in very difference environments. 

\subsubsection{The equation of state of  moderately dense gas}

The value of the Jeans mass is a natural first approach to explain the characteristic mass.
The Jeans mass is a simple function of temperature and density. For a gas described with a barotropic equation of state (EOS), one can calculate the Jeans mass for all densities. 
\citet{Larson85} proposed that there exists a critical Jeans mass that corresponds to the knee in the EOS near $10^5~{\rm cm}^{-3}$ , where the polytropic index goes from $\sim 0.7$ to unity. This critical mass happens around the observed IMF peak and was suggested to be its origin \citep[see also][]{Elmegreen+08}. 
The Jeans mass is expressed as
\begin{eqnarray}\label{eq:jeansmass}
M_{\rm Jeans} &=& {\pi \over 6} {c_{\rm s}^3 \over \sqrt{G^3 \rho}} = {\pi  \over 6} \left( {k_{\rm B} \over \mu m_{\rm p} G}\right)^{3/2} \rho^{-1/2} T^{3/2} \\
&=& 5.1 \times 10 ^{-4} M_\odot \left( {n \over 10^{10} {\rm cm}^{-3}} \right)^{-1/2} \left({T \over 10 {\rm K}} \right)^{3/2}, \nonumber
\end{eqnarray} 
where $k_{\rm B}$ is the Boltzman constant, $\mu$ the mean molecular weight, $m_{\rm p}$ the proton mass, and $n$ the molecular number density. 

\begin{figure}[]
\centering
\includegraphics[trim=0 0 0 0,clip,width=0.8\textwidth]{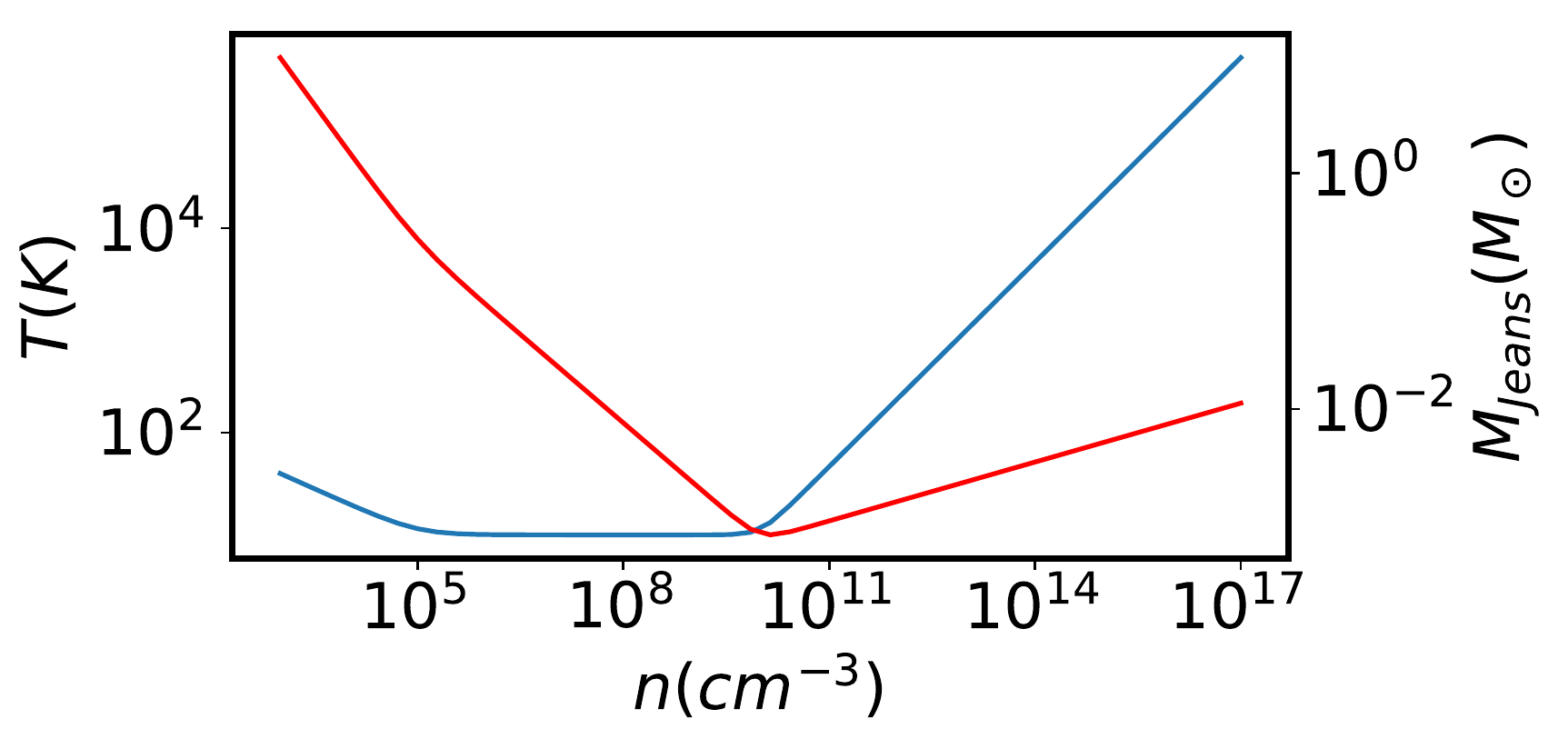}
\caption{Jeans mass (orange) and temperature (blue) as a function of gas number density for an equation of state with polytropic index $\gamma = 0.7$ for $n < 10^5 ~{\rm cm}^{-3}$, $\gamma = 1$ for $10^5 ~{\rm cm}^{-3} < n < 10^{10} ~{\rm cm}^{-3}$, and $\gamma = 5/3$ for $n > 10^{10} ~{\rm cm}^{-3}$. }
\label{fig:Mjeans}
\end{figure}

Fig. \ref{fig:Mjeans} shows the temperature and corresponding Jeans mass for a gas with
the following EOS:
\begin{eqnarray}
P \propto \left\{ \begin{array}{l c l} n^{0.7} & {\rm for }~~ & n < 10^5~ {\rm cm}^{-3}\\
 n & {\rm for }~~ &10^5 ~{\rm cm}^{-3} < n < 10^{10}~ {\rm cm}^{-3} \\
  n^{5/3} & {\rm for } ~~& n > 10^{10}~ {\rm cm}^{-3} \end{array} \right.
\end{eqnarray}
The transition at $n= 10^5~ {\rm cm}^{-3}$ gives a knee in the Jeans mass-density relation around $0.1~M_\odot$. Nonetheless, this is not a local extremum, and thus does not lead to a characteristic mass. On the other hand, the other transition at $10^{10}~ {\rm cm}^{-3}$ actually results in a minimum Jeans mass at several $10^{-4}~M_\odot$, which is much smaller than the IMF peak mass. Consequently, simple arguments from Jeans fragmentation alone cannot easily explain the observed IMF characteristic mass. 
More accurate calculation of detailed description of the thermodynamics show that the minimum Jeans mass corresponds to the observed peak of IMF \citep[][and see also \S~\ref{sec:peak_larson}]{MasunagaInutsuka1999}. 

\subsubsection{The influence of stellar feedback}
A second possible explanation for the characteristic mass is still linked to the Jeans mass of the gas but also accounts for the local effects of stellar feedback \citep{Offner+2009, Bate09b, Krumholz+11, Guszejnov+16}. When a protostellar core forms, it heats the surrounding gas and therefore increases the local Jeans mass, counterbalancing the effect of the increased local density. For example, \citet{Bate09b} derived the effective local Jeans mass when the accretion luminosity heating is taken into account:
\begin{eqnarray}
M_{\rm eff} \approx 0.5~ M_\odot \left( {\rho \over 1.2 \times 10^{-19} {\rm g~ cm}^{-3} }\right)^{-1/5} \left( {L_\ast \over 150 L_\odot}. \right)^{3/10} .
\end{eqnarray}
The effective fragmentation mass near a protostar is therefore close to the IMF characteristic mass and exhibits only depends weakly on the local density and the accretion luminosity. 
However, the physical extent to which heating influences the stellar surroundings depends on a variety of local factors. For low-mass stars, i.e., $M_\ast \simlt 3 M_\odot$ and $\dot M \simlt 10^{-5} M_\odot$yr$^{-1}$, heating extends only to a few hundred to a few thousand au and the temperature distribution depends on the accretion rate, local density, and any gas asymmetries such as the outflow.

\subsubsection{The role of turbulence}
The third argument for the characteristic mass arises from CMF theory, which leads to a modified Jeans mass when density fluctuations are generated in a turbulent medium, as opposed to the conventional Jeans mass, which is derived for a uniform medium. The gravo-turbulent model of \citet{hc08} predicts a CMF peak mass (and the IMF peak by a factor $\sim 1/3$) that is given by
\begin{eqnarray}
M_{\rm peak} = {M_{\rm Jeans} \over 1+ b^2 \mathcal{M}^2},
\end{eqnarray}
where $b$ is a factor around 0.5 that depends on the compressional and solenoidal fraction of the turbulence, and $\mathcal{M}$ is the turbulent Mach number. 
For clouds in virial equilibrium, turbulence balances against self-gravity, giving the relation 
\begin{eqnarray}\label{eq:machnumber}
\mathcal{M}^2 \propto v_{\rm tur}^2/ T_{\rm cld} \propto M_{\rm cld}/(R_{\rm cld}T_{\rm cld}) \propto \rho R_{\rm cld}^2 / T_{\rm cld},
\end{eqnarray}
where the subscript cld indicate the global quantities of the cloud. 
Combining Eqs. (\ref{eq:machnumber}) and (\ref{eq:jeansmass}), the IMF peak mass in a cloud with supersonic turbulence behaves as
\begin{eqnarray}
M_{\rm peak}  \propto \rho_{\rm cld}^{-3/2} R_{\rm cld}^{-2} T_{\rm cld}^{5/2}.
\end{eqnarray}
Taking Larson's relations \citep{larson1981} $\rho \propto R^{-1}$ for the density-size relation and a polytropic equation of state for the diffuse molecular cloud $T\propto \rho^{-0.3}$, 
the peak mass depends on the cloud density, or mass, as
\begin{eqnarray}
M_{\rm peak}  \propto \rho_{\rm cld}^{-0.25} \propto M_{\rm cld}^{0.125}, 
\end{eqnarray}
which is insensitive to the parent cloud mass that can vary by several orders of magnitude. 
The powerlaw exponent depends on the exact scaling relations that are used, but should be a fairly good estimation.  

\subsubsection{The role of filaments}
As discussed in \S~\ref{sec:CMF_obs} and \S~\ref{sec:CMF_theo_filament}, transcritical filaments ($\sim 16~ M_\odot {\rm pc}^{-1}$) are 
common in filamentary star-forming regions. With a typical width of 0.1 pc \citep{Arzoumanian+11}, the gravitational instability that develops at a few times the filament width gives typical fragment masses of a few Solar masses \citep[see e.g.][]{InutsukaMiyama1992, Inutsuka+97}. 
If the filament only fragments along its axis, critical filaments would need to slightly contract in the radial direction before fragmenting, such that the characteristic mass matches with the observed value. 
Meanwhile, the typical Bonnor-Ebert mass expected in such filaments 
is $\sim 0.5~ M_\odot$ \citep[see][and Sect. 5.6 in Chap. 7]{Andre+14, Andre+19}. 
When sub-fragmentation happens inside the {\it groups} (that form from 1D fragmentation), the typical CMF mass is readily recovered.

\subsubsection{The mass of the first hydrostatic core and tidal forces}\label{sec:peak_larson}
Finally, another proposition for the peak mass evokes a natural characteristic mass that arises during the star formation process. 
When the gas number density reaches $n \sim 10^{10} {\rm cm}^{-3}$, the dust opacity becomes significant and its coupling with the gas makes the latter behave adiabatically under compressional heating. 
The collapse of a diffuse isothermal gas is therefore halted at this density, forming a first hydrostatic core, or first Larson core (FLC) \citep{Larson69}. During this phase, the gas is thermally supported and accumulates a mass of $\sim 10^{-2}~M_\odot$before reaching the hydrogen dissociation temperature ($\sim 2000$ K) at the center and triggering the second collapse leading to the formation of a second core or a protostar  (analytically derived by \citet{MasunagaInutsuka1999} as function of temperature of the cloud and dust opacity, or numerically simulated by \citet{Vaytet17} ). Except for stars with very low or high mass, most star formation goes through the above-described process, indicating that the mass of the FLC is a natural lower mass limit for stars \citep[e.g.][]{Bhandare+2018}.

\begin{figure}[]
\centering
\includegraphics[trim=0 0 0 0,clip,width=\textwidth]{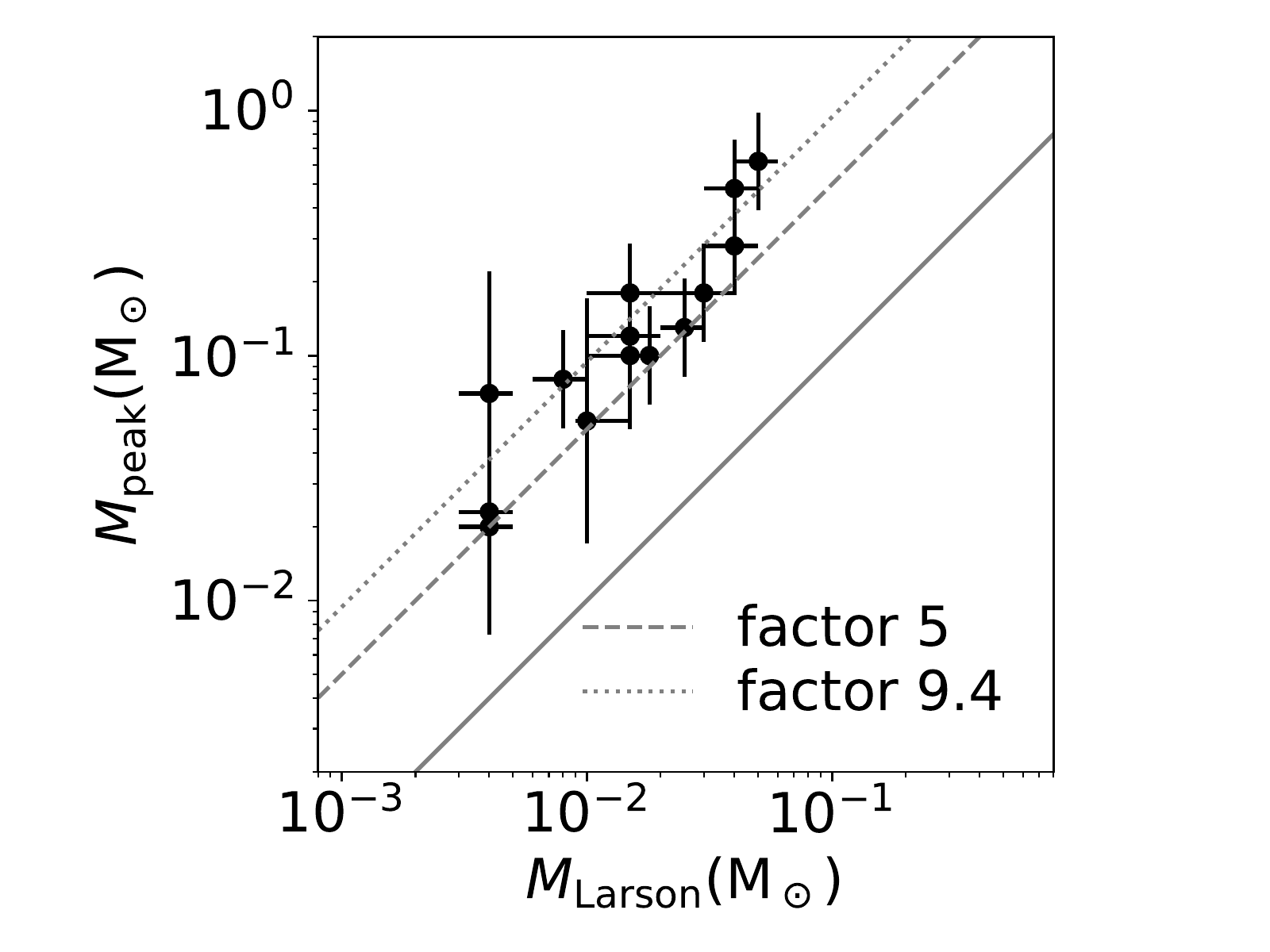}
\caption{The mass of the FLC varies with the choice of the barotropic EOS. Simulations of cluster formation show a strong correlation between the peak of the mass spectrum and the mass of the FLC derived analytically, with a factor of 10 between the two values. Figure adapted from \citet{lee2018b}.}
\label{fig:Mlarson_Mpeak}
\end{figure}

\citet{lee2018b} found a tight correlation between the peak of the stellar mass spectrum and the mass of the FLC derived from the  barotropic EOS, by artificially varying the analytical description of the EOS, with a factor 10 between the two values (cf. \ref{fig:Mlarson_Mpeak}). 
When a FLC forms, this concentrated mass generates a gravitational tidal field around it. This field is destructive in the radial direction and makes any nearby self-gravitating collapse more difficult, thus decreasing fragmentation in its surrounding. 
The non-fragmented gas thus has a chance to be accreted. 
Combining the effect of turbulence, which creates potentially self-gravitating density fluctuations, and the effect of tidal protection around an existing core (protostar), the star that forms within this FLC will eventually reach a final mass ten times its initial mass, that is, $\sim 0.2~M_\odot$, which is very close to the observed IMF peak. \\

\subsection{The high-mass end of the IMF: a powerlaw}

There has been a long standing debate about whether the stellar mass is determined by nature or nurture, that is, whether a protostar, once formed, accretes from a pre-determined mass reservoir or stochastically from the ambient medium. 
Below we discuss two theoretical models that explain the power-law behavior of the IMF from  opposite sides of the nature versus nurture star-formation debate.

\subsubsection{The IMF from stochastic/competitive accretion}\label{sec:BHL}

We first discuss the IMF as derived from stochastic processes. The stochastic mass accretion rate is both related to the medium within which the protostar situates and to the properties of the locally concentrated gas, which is described by Bondi-Hoyle-Littleton accretion \citep{Zinnecker82, Bonnell+01, Edgar04}:
\begin{eqnarray}
\dot{M}  = {4\pi G^2 M^2 \rho_0 \over \left(c_{\rm s}^2 + \delta v^2 \right)^{3/2} }\equiv \alpha M^2,
\end{eqnarray}
where $\rho_0$ is the ambient gas density, $c_{\rm s}$ the thermal sound speed, and $\delta v$ the relative ambient gas velocity. 
With this accretion rate, one can relate the mass of a star to that of the initial stellar seed:
\begin{eqnarray}
M_0 = {M \over 1+\alpha Mt}.
\end{eqnarray}
Therefore, from any initial mass function $\zeta_0(M_0)$ for a population of stellar seeds, the mass function at later time becomes
\begin{eqnarray}
\mathcal{N}=\frac{dN(M)}{d\log M} = \mathcal{N}_0 \left({M \over 1+\alpha Mt}\right) \left(1+\alpha M t \right)^{-1}, 
\end{eqnarray}
which asymptotically approaches $M^{-1}$ for late times. This scenario is examined by \citet{BallesterosParedes+15, Kuznetsova+17, Kuznetsova+18}. Assuming a very simplified physical setup of isothermal gas to avoid other complex effects, they indeed obtained a powerlaw distribution for sink particle masses with $\mathcal{N}(M) \propto M^{-1}$. 
They examined the relation between the instantaneous mass of the sink particles and the corresponding accretion rate and found such relation to be broadly consistent with  Bondi-Hoyle-Littleton accretion, that is, $\dot{M} \propto M^2$. 
\citet{BallesterosParedes+15} showed that this is valid for sinks with similar values of $\alpha \propto \rho/(c_{\rm s}^2 + \delta v^2)^{3/2}$ and that a similar relation holds even if $\alpha$ varies in time, i.e., $\alpha = \alpha(t)$. We note that there is a significant dispersion around this suggested relation, which is probably because estimating $\alpha$ is complicated when the density and velocity structure is complex. 

\subsubsection{The IMF from a pre-determined mass reservoir}\label{sec:IMF_slope_reservoir}

This family of IMF theories are based on the assumption that the mass of a prestellar core forms a protostar with some typical efficiency. Therefore, a theory for the CMF leads directly to an IMF  prediction, sometimes with some further mapping assumptions. 
The CMF models are reviewed in \ref{sec:CMF_theory}, and thus we do not further discuss here. 

\begin{figure}[]
\centering
\includegraphics[trim=0 0 0 0,clip,width=.49\textwidth]{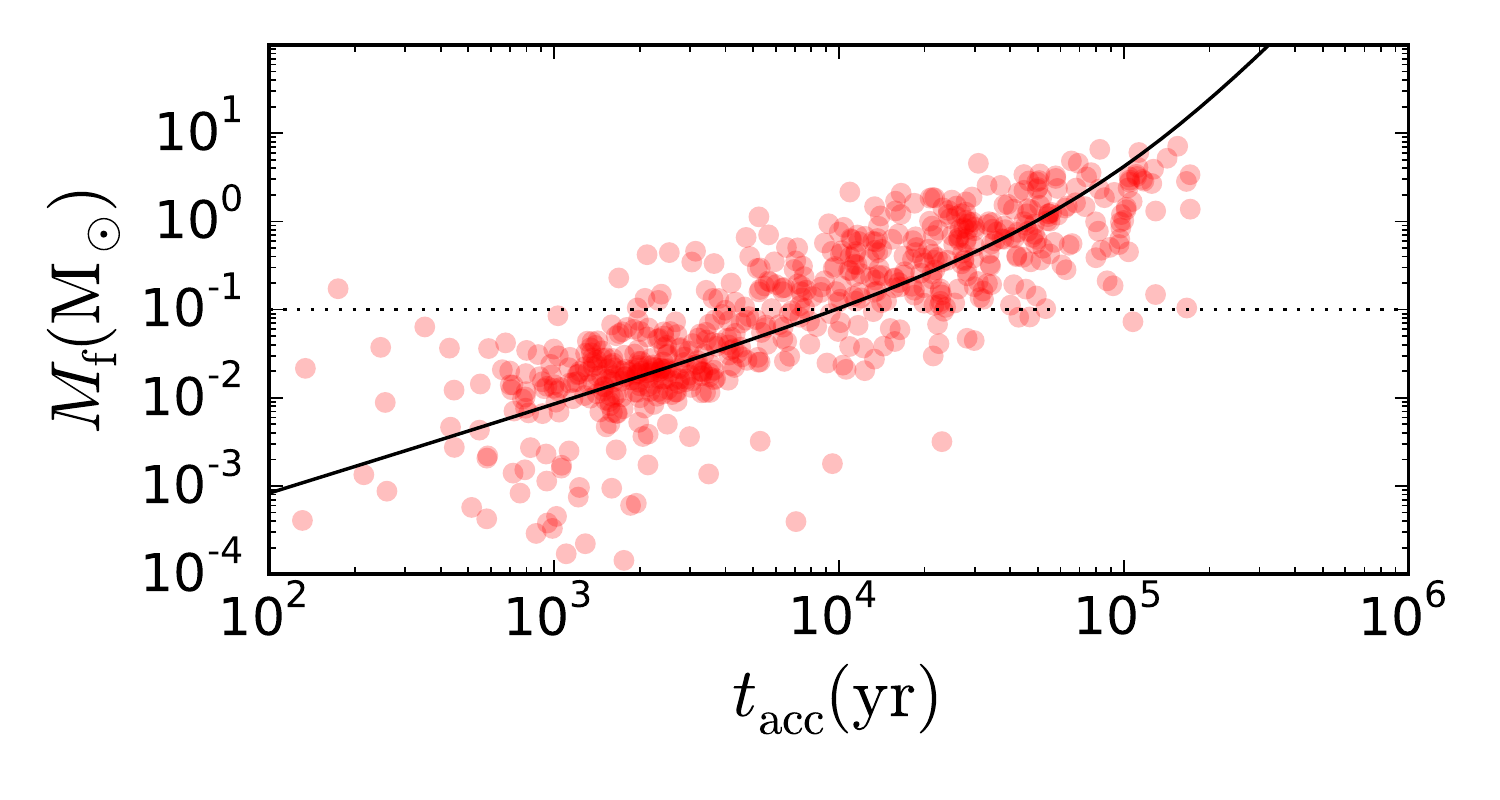}
\includegraphics[trim=0 0 0 0,clip,width=.49\textwidth]{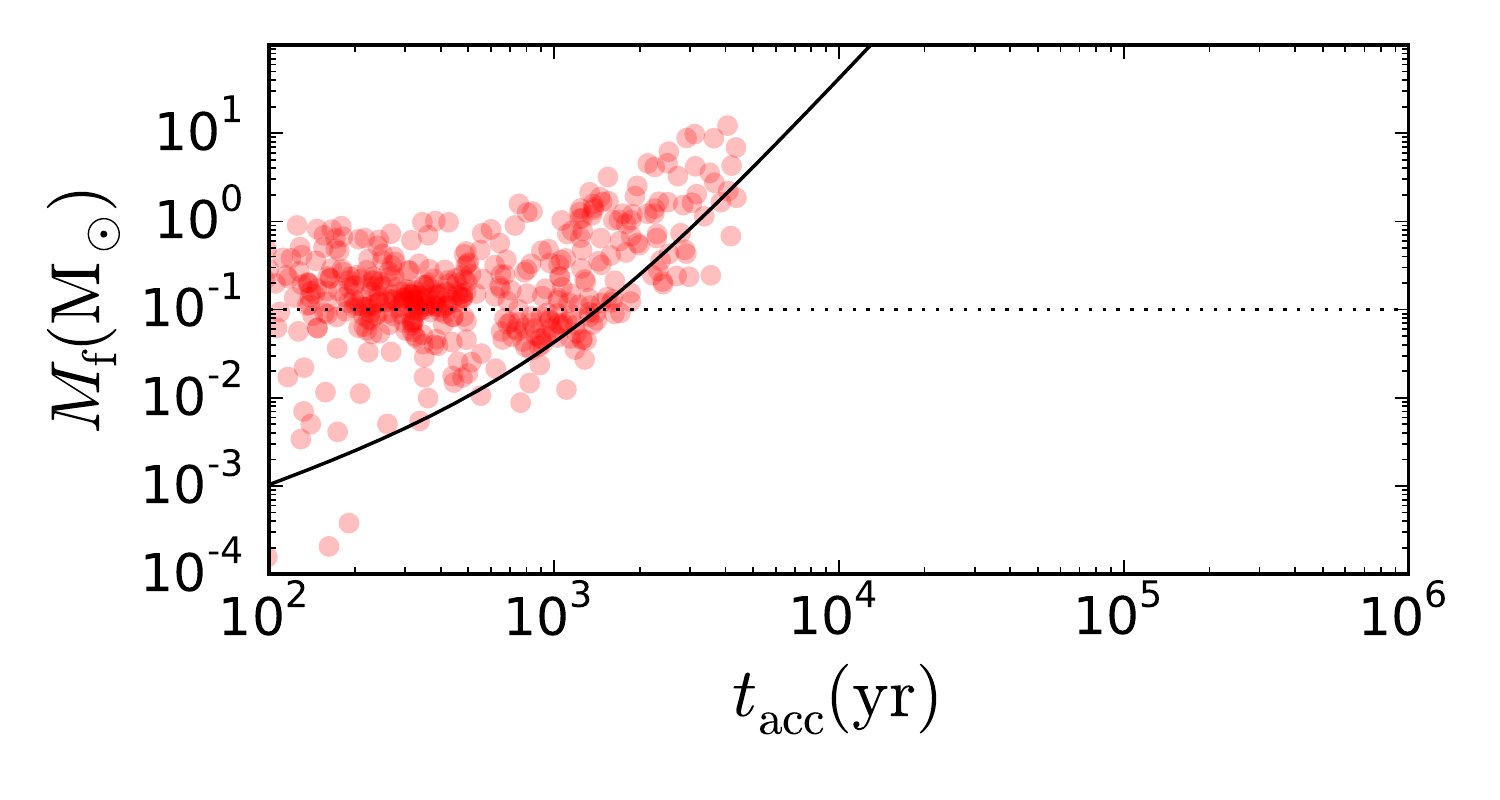}
\caption{Sink accretion time versus mass for two simulations of initially diffuse (left) and dense (right) molecular clouds. The time it takes for a sink to reach $90\%$ of its final mass is plotted against the final mass with red dots. The black curve is the theoretical prediction of the free-fall time of the mass reservoir. Figure extracted from \citet{lee2018a}.}
\label{fig:accretion_time}
\end{figure}

The simplest mapping between the CMF and the IMF assumes a constant $40 \% $ formation efficiency, as suggested by the observed offset between the CMF and IMF peaks \citep[see e.g.][]{kony2015}. In this case, the IMF has exactly the same shape as the CMF, only shifted to lower masses.
\citet{lee2018a} used the CMF model by \citet{hc08} to explain the sink particle mass spectrum and showed consistent results between the theory and numerical results. 
Besides comparing the sink mass function with the CMF prediction, they also compared the time for the sink to reach its final mass and the model predicted for the free-fall time of the mass reservoir, as shown in Fig. \ref{fig:accretion_time}. 
A few conclusions can be drawn from this figure. First, the CMF model does explain the trend of the time-mass relation, although there is a significant dispersion around the model prediction. This means that both a pre-existing mass reservoir and competitive accretion that happens later can play significant roles in determining the final stellar mass. Secondly, the trend is only reproduced above the IMF peak, while below the peak there is a de-correlation between the CMF and the IMF, suggesting different mechanisms for the peak mass as discussed earlier in \S~\ref{sec:IMF_peak}. 

An important unresolved problem is that numerical simulations systematically produce IMF slopes that are shallower than the Salpeter value, mostly shallower than $\Gamma=-1$ \citep[e.g][]{BallesterosParedes+15,Girichidis+11,Krumholz+2012,bonnell2011, Jappsen+05, Bate09b, Bate2012}. For example, \citet{lee2018a} found two regimes of protocluster initial conditions that result in different stellar mass spectra. When turbulence is the main supporting agent against gravity, the mass function behaves as $dN/d\log M \propto M^{-3/4}$. On the other hand, when thermal support is relatively important, $dN/d\log M \propto M^0$. 

As discussed in \S~\ref{sec:CMF_theory}, many CMF theories identify bound structures in a turbulent ISM that is described with a lognormal density PDF. However, \citet{lee2018a} showed that the CMF theory of \citet{hc08} must be adapted with a powerlaw density PDF $\mathcal{P}(\log \rho) \propto \rho^{-1.5}$, which is typical of collapsing gas with density profile $\rho \propto r^{-2}$. Repeating the same exercise, the modified CMF theory recovers the mass function slopes found in simulations. 

Nonetheless, it is important to remember that the entire core mass does not go into the star or even into a bound multiple system \citep[e.g., see discussion of CMF to IMF mapping in][]{Offner+14}. Besides simple mapping with a certain formation efficiency, more sophisticated models consider possible fragmentation during the collapse of the prestellar core. For example, \citet{GuszejnovHopkins15} considered possible fragmentation of prestellar cores using excursion set theory. With larger cores having a higher tendency to fragment, they infer an IMF slope of $\Gamma=-1.3$ and a slightly shallower slope of $\Gamma = -1.1$ for the CMF.

\subsection{The IMF from numerical simulations}\label{sec:IMF_simu}

Simulations aiming to resolve the stellar IMF directly face more extreme numerical challenges than those studying the CMF, since they must model pc scales down to sub-au scales. 
The large dynamic range is usually bridged using sinks particles. Implemented in both smooth particle hydrodynamics (SPH, Lagrangian) codes and Grid-based (Eulerian) codes, sink particles represent the highly concentrated mass that is not resolved with the fluid description. Sink particles are point masses that interact with the rest of simulation only through gravity. Though physically small in size, sink particles are associated with an accretion radius, which is usually a few times the resolution element. Within the accretion radius, the accretion of mass and angular momentum onto the sink particle is assigned numerically with an algorithm of choice. Many codes use a density thresholding algorithm for the sink accretion, putting the excess mass over some density threshold in surrounding cells onto the sink particle. Other codes include more sophisticated criteria such as requiring the total energy to be negative or the flow to be converging  \citep[e.g.,][]{Bate+95,Bleuler14,Federrath2013}.
One caveat is that sink particles are essentially unresolved mass, which can represent anything, according to the physical resolution of the simulation. Only when the resolution is carefully chosen such that the sink particles correspond to individual stars, is it meaningful to compare the mass function of the sinks with the IMF. 

Numerous simulations have been performed to investigate the origin of the IMF
\citep[e.g.,][]{Offner+14}. Numerical simulations that investigate cluster formation are usually initialized with an isolated molecular cloud that is prone to global collapse or a piece of molecular cloud with periodic boundary conditions and turbulence produced through forcing in Fourier space. These simulations essentially have very similar setups to those used for studying the CMF and structure formation within molecular clouds. General processes such as gravity, turbulence, magnetic fields, thermal radiation and stellar feedback are included. 
The first two of these are basic and necessary ingredients; often simplifications are made in each study to reduce the complexity of the problem and to focus on particular mechanisms. The magnetic field is neglected in many studies, since it introduces anisotropy and makes the simulation more complicated. Nonetheless, as discussed in \S~\ref{sec:CMF_simu}, it does not have a strong impact on prestellar core formation, and thus in turn does not affect the IMF. On the other hand, thermal radiation appears to be critical for matching the observed IMF \citep[e.g.,][]{Offner+14}. However, radiation transport is computationally costly, such that a simple isothermal or barotropic equation of state (EOS) is often used to describe the gas temperature, which is a good approximation for the diffuse ISM in general. 
However, after stars begin to form, stellar feedback occurs and the extra heating makes the thermodynamic behavior near accreting protostars very different from a barotropic description. In this case, the thermal radiation must be calculated self-consistently in order to correctly describe the gas thermodynamics. Next, we discuss two categories of numerical studies of the IMF: those that use an analytical EOS and those that consider thermal radiation and feedback from stars.

\subsubsection{Isothermal and barotropic calculations}

An isothermal simulation with gas temperatures of $10-20$ K, the molecular ISM temperature, is the simplest setup to investigate the parameter space of physical conditions that impact the IMF. Since pressure scales linearly with density, isothermal simulations are actually scale-free and can be rescaled to any temperature or corresponding mass/size. Isothermal simulations are therefore used to study essentially the high-mass and scale-free behavior of the IMF with varying cloud parameters such as global density, thermal virial parameter and turbulence properties \citep[e.g.[]{Bonnell+03,Clark+08, BallesterosParedes+15}.  \citet{Girichidis+11} investigated the effect of the cloud density profile and the composition of the turbulence modes (compressive and solenoidal). They found the IMF to be more top-heavy for clouds with an initial central density concentration, while compressive turbulence caused stars to form more rapidly without changing the IMF shape. \citet{lee2018a} and \citet{Guszejnov+18} also found a top-heavy IMF in clouds that rapidly developed a central peak. \citet{lee2018a} studied the effect of initial density and turbulent support and found the top-heaviness comes from the dominance of global thermal energy over turbulent support. \citet{Guszejnov+18}, on the other hand, concluded that a large thermal virial parameter leads to monolithic collapse.

As noted above an isothermal, collapsing gas is self-similar and has no characteristic scale \citep[e.g., see self-similar solutions by][]{Shu1977, WhitworthSummers85}. Therefore, the numerical resolution scale becomes important at some stage and eventually limits the formation of the smallest fragments. Consequently, an isothermal EOS is only useful for studying the high-mass behavior of the IMF, while the low-mass end, or the turnover mass, cannot be revealed by this kind of physical setup. 
Indeed, when gas reaches a star-forming density (say $\sim 10^{10} {\rm cm}^{-3}$) the isothermal approximation becomes invalid, because dust particles become opaque to thermal radiation and thus heat is trapped instead of radiated away freely. 
This allows the gas to heat adiabatically when compressed, with an adiabatic index of $\gamma=5/3$ initially and $\gamma=7/5$ after the excitation of the rotational levels near $100$ K 
. The increased thermal pressure can locally support against gravitational collapse at the center of a prestellar core, therefore limiting the increase of density and subsequent fragmentation. When hydrogen molecules start to dissociate at $\sim 2000$ K, this endothermic reaction causes the effective $\gamma$ to drop to $\sim$1, allowing the second collapse to proceed and form a protostar. 

Thus, it is clear that the IMF is not scale-free all the way down to small masses, and  some mechanisms limit the fragmentation into small objects. A barotropic EOS is the most inexpensive choice that can reasonably mimic the thermal behavior of dense gas. A barotropic EOS with an adiabatic branch at high densities are used in simulations \citep[e.g.,][]{Offner+08} to study the low-mass and brown dwarf regime of the stellar mass spectrum. 
For example, \citet{Bate03, Bate05a, Bate05b, Bate09a, Bate09c} used $\gamma=7/5$ over a critical density and studied the variation of the characteristic mass with different global cloud conditions. The series of studies by \citet{Bonnell08,smith2009,bonnell2011} employed a multi-segment EOS that spans the $\gamma$ variation in the full density range up to protostellar collapse. They show that the non-isothermal part of the EOS is indeed directly linked to the IMF characteristic mass and is needed to explain the full stellar spectrum \citep[e.g.][]{lee2018b,Hennebelle+2019,ColmanTeyssier2020}.

\subsubsection{Simulations with radiation and radiative feedback}

The thermal behavior inside a dense collapsing core is actually more complicated than a barotropic EOS, since thermal radiation depends on the local density, temperature, metallicity, and directional variation of these parameters. A full radiative transfer calculation is therefore needed to study more precisely the behavior of the gas when it approaches star-forming densities \citep{Offner+2009,Bate09b,urban2010,Bate2012,bate2014,Krumholz+2012}. 

At the scale of prestellar cores, heat comes from energy released through gravitational collapse. However, at the stellar scale, which is not numerically resolved, accretion onto the stellar surface and the luminosity from the star itself also contribute to heating the surroundings. For example, \citet{Offner+2009,urban2010,Krumholz+2012} prescribed radiative feedback from the sink particles, including accretion luminosity, and studied the effect of possible self-regulation in the vicinity of a forming star and inside the stellar cluster. These studies found that stellar radiation feedback has a significant
influence on both the cloud and on the resulting mass spectra. In particular, radiative heating reduced the occurrence of disk instability and thus eliminated the excess production of brown dwarfs previously found in calculations without radiation \citep[e.g.,][]{Bate09a}.  
This is due to the intense radiation emitted during the accretion process. 

\citet{Krumholz+2012} also take into account the winds
that are expected to be generated by young protostars. 
They found that radiation can escape through the wind cavity and therefore that the radiation heating is less important;  this eliminated the ``over-heating problem" in which increasing gas temperatures in massive star-forming regions may shift the IMF to higher masses over time. In this case they infer mass spectra that resemble 
the mass spectra at higher masses obtained when the accretion luminosity is not accounted for \citep[e.g.][]{bate2014} or even obtained with 
a barotropic equation of state \citep[e.g.][]{Hennebelle+2019}.
More massive stars, on the other hand, produce ionizing radiation that can influence a larger spatial extent and possibly disrupt the gas inside the star-forming cluster. For example, \citet{He+2019} included this effect to study the IMF and how it regulates the star formation rate and efficiency.

\section{Stellar Multiplicity}\label{sec:multiplicity}

The topic of stellar multiplicity is closely linked to the IMF. Often, observations fail to resolve close multiples and thus measure the {\it system} IMF rather than the IMF of individual stars. 
For unresolved binaries, the lower-mass companion is not measured in the mass function. This results in an underestimation of the low-mass end of the IMF, which can be significant given the high prevalence of (close) binaries \citep[e.g. see review by][]{KroupaJerabkova2018}. Missing the low-mass components can severely undermine the interpretation of stellar evolution of close binaries. 
In order to correct for the unseen binary component to obtain the {\it stellar} IMF, the knowledge of mass ratios in binary systems and their dynamical evolution, which could vary with environment and the primary stellar mass, is also required.  
Besides, exactly how multiplicity varies with stellar mass in turn informs the relationship between the CMF and IMF, which depends on some efficiency factor. Whether binary (or multiple) formation proceeds in the same way in cores of various mass will determine if there is a self-similar mapping from the CMF to the IMF. The binary statistics the low-mass end of the stellar mass spectrum is also a means to distinguish whether the brown dwarfs form through a different collapse process than the low-mass stars and results in a discontinuity in the IMF \citep{Marks+2015}.  Finally, multiplicity properties appear to vary with environment, thus informing both the star formation process and the evolution of star clusters. 

\subsection{Observations}

\subsubsection{Stellar multiplicity of Main Sequence stars}

Chances are if you look at a given star in the sky
that it is actually a member of a binary or multiple system.
Cecilia Payne Gaposchkin once famously said: ``Two
out of three stars is a binary." The history of binary
star research has been reviewed by \citet{Zinnecker2001}. 
in the introductory chapter on IAU Symp. 200 on
"The formation of binary stars" that took place at
the Astrophysical Institute (Observatory) in Potsdam
where some of the first spectroscopic binary star
observations took place (via photographic plates), which
incidentally led to the discovery of the stationary
Ca H and K lines, and hence the interstellar medium \citep{Hartmann1904}. 
The star in question was delta Ori,
a system of two O7V ($20 ~M_\odot$) stars, the middle star of
Orion's Belt, with an orbital period of about 5 days.
The study of this massive main sequence (MS) star and its
short orbital period strikingly foreshadowed what is
today a common result: most massive stars occur in
very tight binary systems with orbital periods below
10 days \citep[e.g.,][]{ ZinneckerYorke2007,Sana+2012}. Such massive stars
have components that are almost touching each other!

For solar-type stars on the Main Sequence, 
the key modern study is the comprehensive work by 
\citet{DuquennoyMayor1991}, 
which concentrated on a complete sample of G- and K-type
star in the solar neighborhood, some 25 pc around the Sun.
They collected previous visual and spectroscopic binary
observations and complemented these where necessary by their
own radial velocity observations. These authors
found that about $50\%$ of solar-type MS stars live in binary
systems; this figure was later revised downward to $45 \%$
by \citet{Raghavan+2010}. 
These binary fractions were the benchmark
against which subsequent studies (starting in the 1990s)
of pre-Main Sequence and protostellar multiplicity rates
had to be gauged, i.e., the multiplicity percentages at birth.
The solar-type MS values are also the reference points
with respect to which the binary and multiple frequencies
of lower mass stars (M-stars, including brown dwarfs)
and higher mass stars (Herbig Ae/Be stars and O-stars)
must be viewed, as we will discuss in the following sections.

\subsubsection{Binarity among intermediate-mass stars}

The best observational survey of the multiplicity among
intermediate-mass stars is the one by \citet{Kouwenhoven+2005}
who used the ESO 3.6m adaptive optics system at 2.2 $\mu m$ to study 199 young A and late B stars in the ScoCenOB2
association (age 5-15 Myr). The stars were selected
using Hipparcos measurements to ascertain membership \citep{deZeeuw+1999}. The detected companion star
fraction was about $50\%$, but detection bias simulations
suggested a higher fraction. \citet{Kouwenhoven+2007}
made a huge effort to estimate the full
primordial binary population using all the
available observations of visual, spectroscopic, and
astrometric binaries with intermediate-mass primaries.
The result is that the binary fraction is at least $70\%$, maybe closer to $100\%$. (They do not consider inverse
dynamical populution synthesis, i.e., the possibility
that the ScoCenOB2 association was much denser 
at the time of formation, which is unknown).
The present separation and mass ratio distribution were found
to be power laws, with index $p_a = - 1$ (Oepik's law)
and $p_q = - 0.4$ (preponderance of small mass ratios).
These results can be considered to be the boundary
conditions for hydrodynamical collapse calculations
(with or without magnetic fields). The authors
consider various pairing mechanisms for these
intermediate-mass binaries and in conclusion
exclude random pairing as well as primary-constrained
random pairing. Indeed some sort of correlated
pairing appears to be consistent with the data.
Similar conclusions (no random pairing, which would
predict a much steeper power-law for the mass ratio
distribution) were reached
earlier by \citet{ShatskyTokovinin2002} for
115 B-star in ScoOB2.
\citet{Janson+2013} surveyed the young intermediate-mass stars 
(early F to late A) in the Scorpius-Centaurus region and found a multiplicity fraction of $\sim 60-80 \%$, supporting the trend of increasing multiplicity with primary stellar mass. 
Contrarily, \citet{Elliott+2015} found no  such trend in their SACY (search for associations containing young stars) survey. 

As for binary statistics among Herbig Ae/Be stars ($1.5 - 8 ~M_\odot$)
we refer to studies by \citet[][63 targets]{BouvierCorporon2001},
 \citet[][7 among 49 targets]{Koehler+2008}, 
and the review by \citet{Duchene2015} concluding
that these young intermediate-mass stars have on
average at least one companion per star (with an
apparent deficit in the 1-50 au range). 
If the companions
are lower-mass T Tauri stars with X-ray emission,
this would indeed explain the puzzling observation
that most Herbig Ae/Be stars turned out to be X-ray sources \citep{ZinneckerPreibisch1994}.
Many Herbig stars show infrared excess indicative
of circumstellar disks, however the companions typically do not. 
The interplay
between multiplicity and circumstellar (protoplanetary) disks
seems similar to that among the lower-mass T Tauri stars.

\subsubsection{Pre-Main Sequence binary and multiple systems}

The statistics of Main Sequence solar-type binaries do not fully reveal
the origin of binary systems, as there are many dynamical
processes between the time of formation (Myr) and the time
when MS binaries and multiple systems are
observed in the solar neighborhood (Gyr later).
Therefore observational studies of young binary systems
are a must, and here we briefly review the results of
some seminal papers of pre-Main Sequence binaries,
the progenitors of those MS G- and K-dwarfs.

In 1993, three independent seminal papers were published
in the span of a few weeks, an indication that the time
and the observational tools to study binary star formation
had come to the fore.
The first study was the 1\arcsec resolution CCD imaging survey at 0.9 $\mu$m
by \citet{ReipurthZinnecker1993} at the ESO-NTT of more than
200 T Tauri stars in nearby southern star-forming regions,
covering projected separations between 150 and 1800 au,
followed by the \citet{Leinert+1993} subarcsec (0.13\arcsec) 2.2 $\mu$m speckle
interferometric imaging survey of some 100 T Tauri stars
in the Taurus-Auriga region together with a similar (0.07\arcsec resolution)
speckle imaging survey of some 70 pre-MS stars
in the Taurus and Rho Oph dark cloud by \citet{Ghez+1993}.
Interestingly, \citet{Leinert+1993} and \citet{Ghez+1993}
differ in one of their conclusions: \citet{Leinert+1993} found
indistinguishable binary fractions among classical (CTTS) and
weak-line (WTTS) T Tauri stars, while \citet{Ghez+1993} found a
significant difference which they attribute to different
(faster) disk evolution for tighter visual binaries, so binaries
with WTTS primaries would be more peaked at closer separations.

A follow-up CCD seeing-limited study to \citet{ReipurthZinnecker1993}
by \citet{Brandner+1996} was exclusively based on X-ray (ROSAT)
selected weak-line T Tauri stars in Taurus and included the
the low-mass young stellar objects in the Sco-Cen OB
association for a total sample size of 195 objects.
High spatial resolution ($\sim~20$ au) speckle studies
found a very high frequency of binary companions
compared to the solar-type and M-type MS numbers \citep{DuquennoyMayor1991, FischerMarcy1992},
in the same range (16-250 au) of separation
(a significant overabundance by a factor of 2-4),
approaching an inferred total frequency of $100 \%$. However, when extrapolating to the full range of semi-major axes,
the two seeing-limited CCD surveys
found a smaller (up to a factor of 2) or no overabundance,
in their observed range of separations (120-1800 au),
consistent with a binary frequency 
slightly higher to that of 
solar-type field stars ($80\%$ vs. $50\%$).

The mass ratios
could also be inferred from component infrared photometry and an
almost linear relation between pre-MS luminosity (flux) ratios
and mass ratios (age-independent for stars on the Hayashi track)
provided the components are coeval.
In general, however, the relation is age-dependent and will also
be contaminated by any active disk contribution to the infrared stellar
luminosity \citep[see discussion by][]{ReipurthZinnecker1993}.
The observed trend shows that companions to the bright T Tauri stars
usually are not near equal brightness (or mass), but rather unequal,
perhaps consistent with random pairing from a low-mass IMF or
with disk fragmentation.
Later, \citet{Correia+2006}, using subarcsec (0.1\arcsec)
adaptive optics techniques at the VLT, showed that
about 1/4 of the primary stars of wide ($> 2$\arcsec ~or $> 300$ au)
T Tauri binaries are themselves resolved close double stars,
suggesting that a fair fraction of young binaries are
actually hierarchical triple (and sometimes quadruple) systems.
This may suggest that multiple systems are prevalent
among young stars and that these systems decay
with time on their way to the Main Sequence
\citep[see discussion by][]{Ghez+1993}. 
For an update on $\rho$ Ophiuchi pre-MS binaries, see \citet[][]{Ratzka+2005}. 
More recent studies by \citet[][$\beta$-Pictoris moving group]{ElliottBayo2016} and \citet[][Taurus]{Joncour+2017} suggest that the majority of very wide binaries ($>1000$ au) have primordial origin and form as a consequence of the structure fragmentation of the natal cloud. These wide systems are usually hierarchical multiples.

Another idea that deserves investigation is that
the population of nearby T Tauri
stars with excess binarity ($100\%$) is not representative
for the majority of progenitor field stars.
It is likely that young clusters and young OB associations
are the dominant birth places of the field star population \citep{MillerScalo1978}.
Thus, the dilemma of excess binaries
among young stars in nearby low-mass star forming regions
would be resolved by a lower binarity fraction
in young clusters. If most young stars originate in clusters,
and if these clusters dissolve and their members become
fields stars, this would greatly alleviate the dilemma
of the above overabundance of binaries compared to the field.
However, studies by \citet{King+2012} look at seven nearby young star-forming regions and finds binary fraction to be decreasing with the stellar density. In particular, there is a distinct excess of closest young binaries (19-100 au) compared to field stars. 
\citet{Duchene+2018} also find twice ($20\%$) the number of stars with close companions (10 - 60 au) in the Orion nebula Cluster (ONC) relative to the field and suggest the multiple system formation mechanisms to be different. 
Moreover, \citet{TokovininBriceno2020} finds a preference for larger mass ratio $q$, i.e., similar mass, at small separations ($\simlt 100$ 100) for early M- and solar-type stars in the Upper Scorpius (USco) star formation region compared to the field. 
This also suggests for some fundamental differences during the formation of these young stars and the field stars.
Future studies of binary statistics of young member stars in
OB associations can be expected after the GAIA data releases.
Much remains to be done 
to investigate
the frequency of spectroscopic binaries among pre-MS stars.
The only semi-systematic study is the one by \citet{Melo2003}
whose result was a fraction of about $10\%$,
but his work is highly incomplete (in terms of epochs
covered) and none has followed it up in the last 15 years.

\subsubsection{Binarity among low-mass stars and brown dwarfs}

Low-mass binaries might form through very different mechanisms and evolve differently than Solar-type binaries. 
Indeed, when the young and dense Orion Nebula Cluster was extensively studied with speckle techniques by \citet{Koehler+2006} (see also \citet{Prosser+1994, Petr+1998}), a pre-MS binary frequency of young low-mass cluster members was found to be lower than that of Solar-type stars.
Note that the low-mass pre-MS stars in the young cluster IC348
also show no excess in binary systems \citep{Duchene+1999}.

The AstraLux survey of M-dwarfs \citep{Janson+2012, Janson+2014} measured a multiple fraction slightly lower than $30\%$, consistent with previous findings \citep[][$32\%$ for 108 M-dwarfs within 52 pc]{Bergfors+2010}.
They also showed a narrower binary separation distribution than Solar-type stars, potentially indicating a continuous decrease toward that of brown dwarfs. 
The M-dwarfs in Multiples survey \citep[MinMS;][]{WardDuong+2015} also found companion fraction of $23.5\%$. 
\citet{Winters+2019} found similar multiplicity fraction ($26.8\%$) and companion fraction ($32.4\%$) in a survey of 1120 M-dwarfs up to 25 pc, leading to $11\%$ of M-dwarf mass hidden in unresolved companions. 
A tendency for separations to be smaller for primaries of lower masses was also suggested by several studies \citep{Kraus+11,WardDuong+2015,Winters+2019}. 

Dynamical studies of very low mass binaries  \citep[VLMBs][]{ParkerGoodwin2011} showed that those with small separations ($<20$ au) are extremely hard to destroy and can reflect their birth conditions, they are therefore useful for comparing young star-forming regions with the galactic field. 
Binary fractions of low-mass stars in dense star-forming regions resemble that in the field \citep[e.g.][ONC]{DeFurio+2019}, while an excess is found in low-density clusters \citep[][Taurus]{Kraus+11}.
In several young associations, \citet{KrausHillenbrand2012} found the same trend of decreasing binary fraction and decreasing separation for lower-mass binaries as in the field. 
\citet[][]{Todorov+2014} also found decreasing binary fractions from $15\%$ for M4-M6 dwarfs to $4\%$ for dwarfs later than M6 in young star-forming regions (Taurus, Chamaeleon I, and Upper Sco). 
They found more wide binaries than in the field, suggesting that, such binaries were inhibited from forming in the natal condition of the field stars or they have been disrupted on a time scale longer than the life time of these young regions. 

\citet{ThiesKroupa07} suggested a discontinuity in the IMF near the hydrogen-burning limit due to distinct formation pathways. Brown dwarfs tend to from from fragmentation of circumstellar disks, and low-mass stars collapse directly from a prestellar core, whereas these two channels might not be mutually exclusive. 
The statistics of low-mass binaries is therefore a useful way to probe such differences. 
For example, significant difference is found in the mass ratio of brown dwarf binaries, compared to the flat distribution of that of stellar binaries \citep{Goodwin2013,Fontanive+2018}. 
Low binary fraction is suggested both in the field \citep[$<10\%$][]{Fontanive+2018}, in Pleiades \citep[$<11\%$][]{Garcia+2015}, and in Upper Sco \citep[$<9\%$][]{Biller+2011}, which is consistent with a continuous decrease with spectral types.

Subdwarfs in the galactic halo are another group of interesting objects to study, which can provide hint on the effect of low-metallicity star-forming environments. 
\citet{Jao+2009} found multiplicity rate of $26 \pm 6 \%$ for 62 cool subdwarf systems within 60 pc, which is lower than that of their MS counterparts ($37 \pm 5 \%$). 
This value was decreased to $\sim 10\%$ later by \citet[][SDSS,1800 objects]{Zhang+2013}
and \citet[][344 objects]{Ziegler+2015}.
All these studies support trends for wider binary separation (mostly larger than 100 au) than the K-/M-dwarfs, and decreasing binary fraction with decreasing mass and metallicity.

\subsubsection{Protobinaries (Class I and Class 0 young stellar objects)}\label{protobin}

       The search and characterization of protobinary stars
       extends the observations of T Tauri pre-MS binaries to
       the youngest stages, when the components acquire the bulk
       of their stellar mass, either from envelopes or massive
       stellar disks. Stars deeply embedded, not detected in the
       near- or even mid-infrared, but only in the far-infrared
       and submillimeter domain, are denoted Class 0 objects (true protostars)
       totally obscured with massive disks and vigorous jets/outflows.
       By contrast, Class I objects are less extreme, visible in the
       near- and thermal infrared and a slightly more evolved stage
       of accreting protostars \citep[][]{Andre+93}. 
       In this scheme, classical T Tauri stars with accretion disks
       correspond to Class II objects, while weak line T Tauri stars
       with weaker or no circumstellar disks are referred to as Class III.
       There is an evolutionary sequence from Class 0, I, II, to III
       with associated timescales of $10^4,~ 10^5, ~10^6{\rm ,~ and}~ 10^7$ years
       respectively (roughly speaking) \citep{Dunham+2014}. 

 One fundamental goal of studying the youngest multiple systems is to characterize their gas environment and thereby identify clues about their origin. Class 0 companions are of particular interest since they are considered too young to have migrated from their birth positions.  To date, a number of wide-separation ($>10^3$ au) Class 0 sources have been identified \citep{Maury+10,Chen+2013,LeeLee+2017}. One high-resolution study discovered a {\it quadruple} system forming within a filament, which includes  three gravitationally bound gas condensations that do not yet contain protostars \citep{Pineda+2015}.  Study of the surrounding gas properties of such systems suggests they are created by ``turbulent" or filament fragmentation of the parent core (see \ref{sec:corefrag}). Meanwhile, two additional protostellar systems, a binary and a triple, exhibit separations $<100$ au and are embedded within a large, shared disk \citep{Tobin+2016,Alves+2019}. These systems are likely produced by disk fragmentation (see \ref{sec:disk}).   

     A variety of surveys have targeted protostars in nearby star-forming regions in order to obtain a more complete statistical picture of the youngest multiple systems \citep{Reipurth+2014}.
      Research into the statistics of protobinary systems started in earnest with a
       large survey by \citet{Connelley+2008, Connelley+2009}.
       The targets are all within 1 kpc and the observations were
       carried out in the L$^\prime$ (3.6 $\mu$m) band at the UH 2.2 m telescope.
       They detected 89 companions of which 73 were new detections.
       They speculated that the subset of close Class I binaries,
       resolved by Subaru and Keck adaptive optics observations
       form via ejection during the early dynamical decay of
       non-hierarchical multiple systems, which may have formed
       by turbulent fragmentation or fragmentation of filaments
       \citep[first discussed by][]{Zinnecker+1987, BonnellBastien1992}.
       Also noteworthy are the surprise observations of ``infrared
       companions'' to T Tauri stars \citep{Chelli+1988, WilkingZinnecker1992}, which are systems where the companion
       becomes brighter at longer infrared wavelengths, perhaps
       indicating components consisting of Class II and Class I
       objects!
       
        Two early interferometry surveys of Class 0 sources concluded that protostellar multiplicity is higher than that of the field population, although they disagree on whether multiplicity increases or decreases from the Class 0 to Class I stage \citep{Maury+10,Chen+2013}.
       Recently, the VANDAM VLA/ALMA survey of protostars in Perseus made the next big step towards discovering new Class 0 and I  protobinaries 
       \citep{Tobin+2018a}. 
       One of the goals of the VANDAM
    survey was to discriminate between binary formation
       processes, such as disk fragmentation, as indicated by aligned binary
       spins, outflows and disks, from turbulent core fragmentation, which produces disks/jets with more random orientations.
       They studied 17 multiple sources (9 Class 0 and 8 Class I) in an
       ALMA 1.3 mm follow-up of their large VLA 7 mm/4 cm survey.
       12 out of the 17 sources were also resolved with
       ALMA (0.27 x 0.16\arcsec beam). In eight out of the 12 cases
       gas velocity information points to disk fragmentation,
       while the other 4 systems are better modelled by
       a variant of turbulent fragmentation. Circumbinary structures
       were detected around the Class 0 sources but not
       around the Class I sources. 
       
       A complementary survey to VANDAM is the MASSES
       SMA Legacy survey 
       of 73 protostars in Perseus, one goal of which is to connect multiplicity with the properties of their outflows and host dense cores \citep{Lee+2015}. MASSES finds that protostellar multiplicity is hierarchical with separations in a given system ranging from $<100$ au to a $>10^3$ au \citep{Lee+2015}. The angle distribution between outflows in wide separation pairs ($>10^3$ au) is consistent with a random distribution, indicating that the angular momentum in these systems is disordered, possibly due to turbulence \citep{Lee+2016}. 
       
       Another approach to understand binary formation involves looking for gas substructure within cores, a predicted signature of turbulent fragmentation \citep{Offner+2012,Mairs+2014}. An ALMA survey by \citet{Dunham+16} of Chamaeleon I detected 56 starless cores, none of which show evidence of turbulent fragmentation. However, this may be because the cores are not gravitationally bound and thus not star-forming. A similar ALMA survey of 60 cores in Ophiuchus detected two starless cores with substructure \citep{Kirk+2017}. 

       In the future, the ngVLA and JWST hold great promise
       to improve protobinary statistics by extending the
       distance-limited samples of protostars to about 1.5 kpc
       and also obtain 15 au resolution in the Orion region \citep{Tobin+2018b}.

\subsubsection{Summary of observed multiplicities as
       as a function of primary mass and environment}

       The main conclusion of the last years of studies
       of young binary stars of all primary masses has
       been that the binary frequency steadily increases
       with mass \citep[see review by][]{DucheneKraus2013}:
       from about $30\%$ for M-type stars ($<0.5~M_\odot$),
       to about $50\%$ for solar mass K- and G-type MS stars
       (but higher for solar-type pre-MS stars)
       up to $100\%$ and more for MS stars of
       spectral type earlier than B2 ($>10~M_\odot$). The most
       massive stars are extremely rarely single, and
       if so, this is likely due to dynamical evolution, e.g., very close massive binaries merge or one component in the binary star system explodes
        as a supernova and kicks away the other member,
        creating a runaway O-star. The most massive
       stars often consist of a close spectroscopic pair
       with orbital separation less than 1 au and a
        wider bound component at distances 100-1000 au
       \citep{Mason+1996}. The jury is still out whether
       the most massive stars in dense clusters have
       different multiplicity properties from those in
       looser OB associations, but barring more detailed
       studies the multiplicity fractions of massive stars
       as a function of environmental factors (including
       stellar density and even metallicity, such as the LMC versus SMC)
       appear to be indistinguishable and very similar \citep[see the review by][]{Sana2017}.

\subsection{Theoretical Models}

A successful model describing multiple formation must be able to explain the wide range of observed separations and the distribution of observed mass ratios. Indeed, this is daunting task for any one model and recent observations of protostars hint that multiple mechanisms governed by different physical processes are at work. 

In this section we consider the three main theories for the origin of stellar multiplicity:  fragmentation of the parent dense core to form bound companions, gravitational fragmentation of an unstable accretion disk and dynamical evolution, in which multiple systems may form via gravitational capture or loose members through interactions. Each mechanism acts on different physical scales and times within the star formation process. These differences may ultimately allow them to be observationally distinguished and explain the diverse nature of observed stellar multiples.

\subsubsection{Core Fragmentation} \label{sec:corefrag}

The theory of core fragmentation, which is also referred to as ``prompt fragmentation" or ``turbulent core fragmentation," has its genesis in the seminal work of \citet{Hoyle1953}. \citet{Hoyle1953} proposed the idea of hierarchical star formation, in which gravitational fragmentation proceeds to increasingly small scales until fragments no longer cool efficiently and thus become thermally supported (opacity limited fragmentation). 
While Hoyle self-deprecatingly described the idea of hierarchical fragmentation as ``of a mainly tentative character" and ``too qualitative" for observational comparison, it inspired numerous additional studies and provided a basis for more modern work on binary formation.

The presence of angular momentum is a fundamental prerequisite of binary formation.  Early idealized theoretical models for star formation described the collapse of dense cores beginning with spherically symmetric, self-similar density and velocity configurations \citep[e.g.,][]{Shu1977}. The addition of rotation naturally causes a spherically symmetric core to fragment into two or more objects \citep{Larson1972,Bonnell1994}. The outcome is dictated by the initial ratio of thermal to gravitational energy, $\alpha = 5c_{\rm s}^2R/GM$, and the ratio of rotational to gravitational energy, $\beta = \Omega^2R^3/3GM$, where $c_{\rm s}$ is the sound speed, $R$ is the core radius, $M$ is the core mass, and $\Omega$ is the core rotational frequency. 
Isothermal, collapsing rotating cores are prone to fragmentation when $\alpha \beta < 0.12$ and $\alpha < 0.5$ \citep{TsuribeIntsuka1999a,TsuribeIntsuka1999b}.

The origin of angular momentum on core scales is thought to be inherited from the larger cloud environment. Numerical studies of turbulent cores show that turbulence naturally generates velocity gradients and provides sufficient angular momentum to produce fragmentation \citep{BurkertBodenheimer2000,Goodwin+2004}.  Semi-analytic models of cores including turbulence are able to reproduce the observed binary distribution from turbulent fragmentation alone \citep{Fisher2004,JumperFisher2013}.

The impact of magnetic fields on the process of multiple formation remains debated.
Classically, magnetic fields are predicted to suppress collapse and fragmentation by providing additional pressure support \citep{Mouschovias1976,MouschoviasSpitzer1976}. Magnetic fields may also efficiently remove angular momentum via magnetic breaking \citep{Mestel1979,Mouschovia1977}. 
Numerical simulations of rotating, spherical magnetized cores demonstrated that strong fields inhibit fragmentation \citep{HoskingWhitworth2004,Fromang+2006,PriceBate2007a}. However, since gas more easily collapses along rather than perpendicular to field lines, magnetic fields introduce a preferred direction for collapse, which produces a bar structure \citep{Dorfi1982,Benz1984,Boss2000}.
This filamentary structure may be unstable to fragmentation thereby enhancing binary formation \citep{InutsukaMiyama1992,Boss2000,Machida+2004,Machida+08}, especially in the presence of turbulence \citep{Offner+2016}. The resulting binary properties reproduce the distribution of mis-aligned outflows observed in wide, binary protostellar systems as well mis-aligned stellar spins \citep{Offner+2016}. These characteristics serve as observational signatures of turbulent core fragmentation.

\begin{figure}
 \includegraphics[width=\linewidth, trim=0.5cm 11cm 0.5cm 0.0cm,clip]{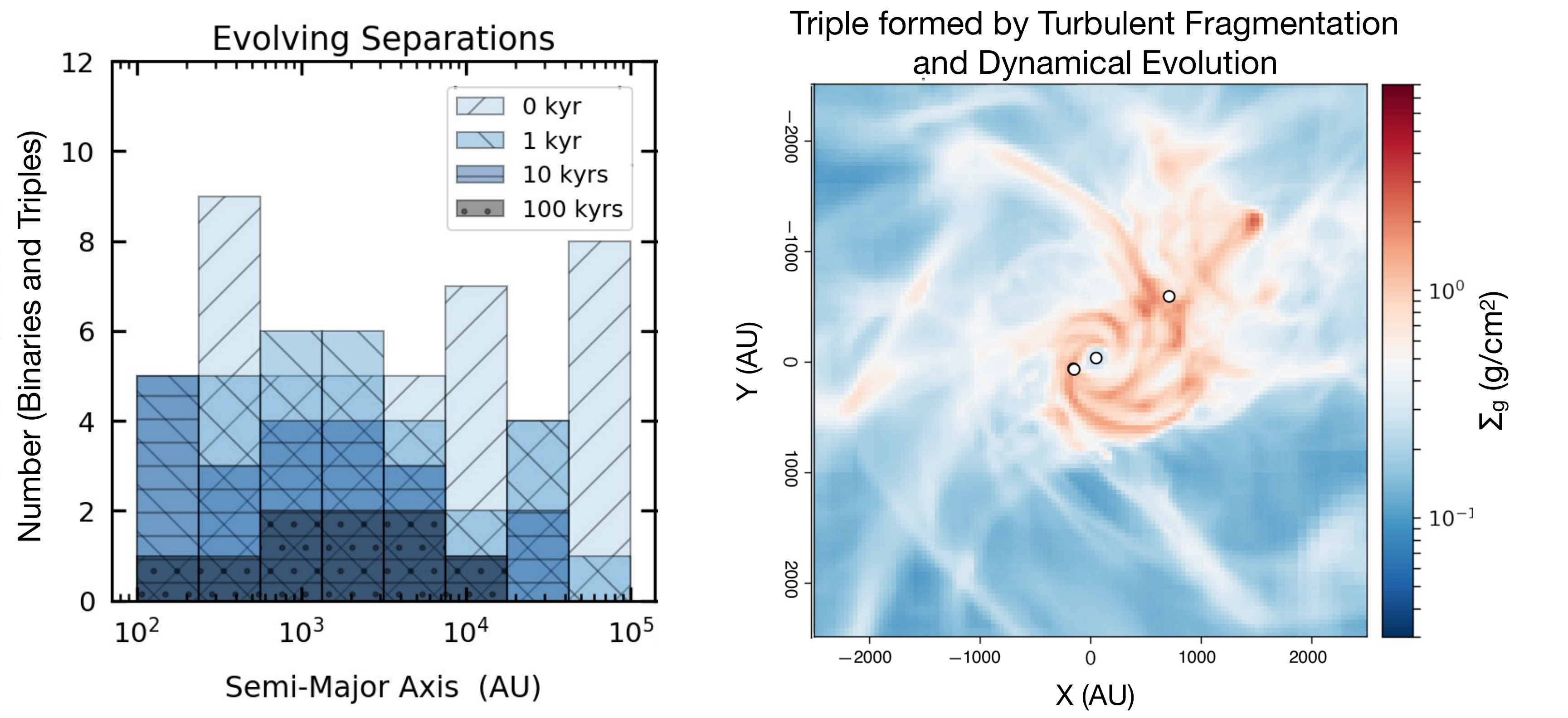} 
 \caption{Left: Distribution of protostellar pair separations for binary and triple systems formed in MHD star-cluster simulations binned by their age. Protostars tend to evolve from wider to closer separations within 100 kyr. Right: Gas column density around a triple protostellar system (circles). Although the disk appears gravitationally unstable, its extended, spiral morphology is a product of the perturbation caused by the migration of the tertiary from its formation location $>$1000 au inwards.
Both panels adapted from \citet{Lee+2019}.}
\label{multiplicity}
\end{figure}

The net result of turbulent core fragmentation is companions with initial separations of a few hundred to a few thousand au \citep{Offner+2010,Bate2012,Offner+2016,Kuffmeier+2019,Lee+2019}. However, protostars formed this way are initially not on stable orbits and they migrate on relatively short timescales ($\sim$0.1 Myr) to $<$100 au~separations or become unbound through dynamical interactions (see \S \ref{sec:dynamic}). Figure \ref{multiplicity} shows the pair separation evolution of a population of binaries and triples formed in a star cluster simulation via core fragmentation; once pairs reach close separations the resulting gas distribution and stellar configuration may resemble systems formed via disk fragmentation (see \S \ref{sec:disk}). We return to the process of core fragmentation acting in concert with other binary formation mechanisms within star cluster environments below.

\subsubsection{Disk Fragmentation}\label{sec:disk}

The formation and evolution of disks play a crucial role in mediating the star formation process \citep{Zhao+2020}. Here we focus on disks as the sites of multiple formation. Sufficiently massive disks are prone to gravitational instability, which seeds the formation of close ($< 200$ au), stellar companions \citep[see][for a recent review on gravitational instabilities]{KratterLodato2016}. The criterion that dictates disk stability is given by 
\begin{eqnarray} 
Q = \frac{c_{\rm s} \kappa}{\pi G \Sigma},
\end{eqnarray}
where $c_{\rm s}$ is the thermal sound speed, $\kappa$ is the epicylic frequency and $\Sigma$ is the surface density \citep{Toomre1964}. As $Q$ decreases self-gravity becomes increasingly important. Spiral structure begins to develop when $Q\sim 2$ and fragmentation occurs when $Q \sim 0.6$ inside the spiral arms \citep{Tsukamoto+15a,Takahashi+16}. Not all fragmentation events produce stellar companions, however.  Some fragments rapidly migrate inwards, merging with the central protostar and triggering accretion bursts \citep{VorobyovBasu2005,VorobyovBasu2010,ChaNayakshin2011}. Fragments may also interact with one another and be dynamically ejected (see \S \ref{sec:dynamic}). Only those fragments that continue to accrete mass and achieve stable orbits will become stellar companions \citep{VorobyovBasu2010,Kratter+2010a,Tsukamoto+13,Tsukamoto+15a}.

One implication of the Toomre criterion is that colder, more massive  disks are more prone to gravitational instability. Numerical simulations show that radiative feedback from the central source heats the disk, significantly reducing the liklihood of fragmentation \citep{Krumholz+2007,Offner+2009,Bate09a,Offner+2010}. However, if angular momentum transport is inihibited in the inner disk, accretion may slow, thereby reducing the impact of radiative feedback and allowing fragmentation to proceed \citep{StamatellosWhitworth2011}. 

Magnetic fields also play an important role in regulating the disk size and efficiency of angular momentum transport. In the ideal MHD limit, simulations show that coupling between the small-scale field near the protostar and the larger-scale core field can facilitate the removal of angular momentum from the collapsing region, potentially preventing a disk from forming, the so called ``magnetic breaking catastrophe" \citep{Galli+2006}.  Consideration of non-ideal effects, turbulence and angular momentum-field misalignment mitigate the impact of magnetic breaking \citep{MouschoviasSpitzer1976,HennebelleCiardi2009,Joos+2012,Krumholz+2013, Tsukamoto+15b,Tsukamoto+15c,Tsukamoto+17,Tsukamoto+18}. Thus, strong magnetic fields may reduce the incidence of disk fragmentation since small disks are less prone to fragmentation.
However, the magnetic field dissipates in the high-density region (dead zone) where magnetic braking does not work and self-gravitational fragmentation of a massive disk frequently occurs \citep{Machida+09, Inutsuka+2010, Machida+12}. Thus, the dissipation of the magnetic field may play a primary role for disk fragmentation \citep[see also][]{Machida+08}. 

Stellar companions produced via disk fragmentation likely form during the main accretion phase ($\simlt 0.2$ Myr), when infall is the highest and the disk is the most massive. Disk fragmentation produces companions with initial separations $\sim 50-200$ au, i.e., in the disk region that is cold while still containing a significant mass reservoir \citep{Kratter+2010b}.
Unstable disks may also produce more than one companion and so provide a mechanism to create higher-order multiple systems \citep{Kratter+2010a,StamatellosWhitworth2011}.

The frequency of disk fragmentation depends on a number of factors, including the rate of infall, core angular momentum, magnetic field properties and stellar heating \citep{matsumoto2003,Kratter+2010a,VorobyovBasu2010,Offner+2010,Inutsuka+2010,Tsukamoto+15a}. 
Numerical simulations demonstrate that cores with high infall rates, such that those that produce massive stars, produce disks with higher rates of instability \citep{KratterMatzner2006,Krumholz+2009,Rosen+2016,Matsushita+2017}.  This effect is qualitatively consistent with the higher multiplicity fraction characteristic of massive stars. However, dynamical evolution is necessary to produce both very close and wide separation companions from the initial separation distribution as discussed below.

\subsubsection{ Dynamical Evolution \& Capture}\label{sec:dynamic}

The final way multiple systems may form is through gravitational capture during dynamical interactions \citep{Bate03,Bonnell+2004,GoodwinKroupa2005}. In this scenario, members of multiple systems are not initially gravitationally bound and did not form within the same over-density. Capture is facilitated through dynamical friction with the gas, the presence of a disk, or n-body interactions involving three or more stars \citep{ClarkePringle1991,MoeckelBally2007,ReipurthMikkola2012}. This mechanism is generally most efficient in dense, clustered environments, where close encounters are frequent \citep[e.g.,][]{MoeckelBate2010}. The likelihood of capture is highest before and during cluster dispersal, after which capture is inefficient in forming new multiples  \citep{MoeckelClarke2011}. N-body simulations of dissolving star clusters predict that the formation of wide binaries and their separation depend on the cluster mass and size, respectively, with the fraction of wide binaries decreasing with cluster mass \citep{Kouwenhoven2010}. The diversity in star cluster initial conditions may help explain variation in the fraction of binary systems having separations $>10^3$ au.

Dynamical evolution also shapes system architectures by modifying stellar separations and reducing the number of members in higher order systems. Angular momentum exchange between stars, gas and cluster members cause closer orbits to shrink to separations less than 50 au and wide orbits to become wider, exceeding separations of $10^3$ au \citep{Bate03,Stahler2010,Offner+2010,Lee+2019}. 
Dynamical interactions of initially compact triple star systems can produce hierarchical triples, where a close pair is orbited by a widely separated tertiary companion \citep{ReipurthMikkola2012}.
Close encounters and n-body interactions may cause members to be completely ejected from the system \citep{ReipurthClarke2001,Bate+2002}. Dynamical interactions also impact secondary multiplicity characteristics such as disk, outflow and stellar spin alignment \citep{Bate2012,Offner+2016}. The rapidness of the evolution, especially during the first 0.1 Myr,  underscores the importance of studying multiplicity during the earliest stages of star formation.

\subsubsection{Multiple Formation in Star Clusters}

A fundamental implication of the above theories is that cores and disks cannot be divorced from their birth environment. Cores often form in dynamic, clustered regions, embedded within filamentary gas flows (see \S \ref{sec:CMF_obs_filament}). In the context of star cluster formation, hydrodynamic simulations suggest that all three of these multiple formation mechanisms operate.

The dominant mode of binary formation is sensitive to the physical processes included. For example, when radiative transfer is included in star cluster calculations, the incidence of disk fragmentation is significantly reduced \citep{Bate09a,Bate2012}. Radiative feedback from forming stars further reduces disk fragmentation and demonstrates that turbulent fragmentation is a viable mechanism for producing binaries in simulations of low-mass ($M< M_\odot$) star formation \citep{Offner+2009,Offner+2010}.
Star cluster simulations that compare different magnetic field strengths show that more magnetized clouds produce comparable or higher binaries fractions than less magnetized clouds, although the overall star formation rate is lower \citep{Cunningham+2018,Lee+2019}. Here, magnetic support rather than radiative feedback may act as the stabilizing mechanism on small scales \citep{Lee+2019,Kuffmeier+2019}.
However, comparison to observations suggests close-separation ($<10$ au) binaries are under-produced by turbulent fragmentation alone.

In principle, observed multiplicity metrics can be used to benchmark numerical calculations, assess the role of different physics and determine realistic initial conditions. However, comparisons are frustrated in part due to the poor statistics in many cluster calculations and in part due to the relatively large uncertainties in the observations. Consequently,
a variety of calculations with a range of different physics produce multiple system properties that  compare favorably with observed statistics. One of the most extensive samples of simulated multiple systems was produced in \citet{Bate2012}, a calculation which included radiative transfer but excluded magnetic fields and outflows. The simulated systems reproduce the observed field star multiplicity fraction, semi-major axis distribution and mass-ratio distributions.
Simulations of magnetized, collapsing clouds with radiative feedback and outflows also reproduce the field multiplicity fraction as a function of primary mass but with less statistical significance \citep{Krumholz+2012,Cunningham+2018}.  
However, hydrodynamic calculations such as these are evolved at most only a few cloud free-fall times, at which point most sources are still protostellar and their multiplicity properties should not necessarily be expected to match the multiplicity of the much older field population (see \S\ref{protobin}).  Both observational and theoretical progress is required for observational comparisons to discriminate between different numerical models and initial conditions.

Fully modelling the formation and  evolution of multiple systems and explaining the multiplicity of field stars is arguably a more challenging theoretical problem than reproducing the stellar IMF alone. It requires following star formation through gas dispersal, including both realistic initial cloud conditions, which are critical for setting the IMF, {\it and} evolving calculations through the gas dispersal phase, which spans tens of Myr. In addition, stellar feedback such as protostellar outflows, winds, radiation and supernovae play important roles in setting cluster efficiencies and dissolution timescales by regulating gas dispersal \citep{Krumholz+2019}. 
These factors altogether circumscribe a physically complex, nonlinear problem that exceeds the limits of current numerical calculations and presents a formidable challenge for future theoretical models.

\section{Conclusions and Outlook}\label{sec:outlook}

In this chapter, we have reviewed the statistical properties of star-forming clusters, more precisely, the mass distributions of dense cores and stars. 
Ever since the first IMF measurement by \citet{Salpeter55}, 
enormous efforts have been directed towards testing the universality of the IMF. With improved observational power, it is now possible to probe star-forming regions and star clusters at larger and larger distances. Some variation is indeed suggested in extreme environments that differ significantly from the Solar Neighborhood.  

At the same time, observational sensitivity and resolution have improved such that  prestellar cores can be readily observed and confident prestellar CMFs can be measured,
enforcing the idea that the IMF might be inherited directly from the CMF. However, there are still some differences between the working definitions of cores in varying studies, in particular between observations and theories, complicating direct comparisons. 
Consequently, the true relationship between the CMF and IMF remains debated.

We describe  several 
theories that try to explain the observed form of the IMF and achieve varying degrees of success.
This reflects the fact that the star formation inside turbulent molecular clouds with multiple physical mechanisms is complex.  
It is possible these theories are not mutually exclusive and describe cluster formation under different conditions. At the same time, numerical simulations are becoming more 
sophisticated and including more realistic physics. 
In contrast to simulations with simplified physics that study the role of one particular mechanism, 
multi-physics simulations provide better understanding of the effects of non-linear coupling between different physical processes in regulating stellar masses.

Although computational speed continues to increase according to Moore's law, important physical effects such as radiative transfer scale poorly to large numbers of cores. Multi-physics calculations including dynamic ranges spanning au to pc scales will continue to be challenging. Ideally, future calculations will also move towards more realistic initial conditions. 
Recent efforts have explored binary formation starting on galactic scales and ``zooming in"  to au scales \citep{Kuffmeier+2019}. However, statistical significance is even harder to achieve in this approach. More efficient algorithms, new methods exploiting GPU capabilities and the application of machine learning to model statistical processes like dynamical interactions show promise for future progress \citep{Schive+2018,Nordlund+2018,Vanelteren+2019}.

Besides the overall statistics of the IMF in clusters, progress has been made in understanding the origin and statistics of binaries and small multiple systems. These studies have important implications for the collapse process of prestellar cores, 
which can often result in fragmentation. Studies of these small systems also allow a better understanding of star formation at the prestellar core and stellar scales. 

In summary, we are in an era where we start to really test the possible variations of the IMF, with tools from many aspects. Important physics of the star formation process can be unveiled by either searching for IMF non-universality or developing models for the IMF that can explain its invariance. 
Whatever the outcome should be, there will be a lot to learn. 

\begin{acknowledgement}
We thank the International Space Science Institute (ISSI) for generously providing such stimulating environment for collaboration. Y.N. Lee acknowledges funding from the Ministry of Science and Technology, Taiwan (grant number MOST 108-2636-M-003-001), the grant for Yushan Young Scholar from the Ministry of Education, Taiwan, and the UnivEarthS Labex program at Sorbonne Paris Cit\'e (ANR-10-LABX-0023 and ANR-11-IDEX-0005-02). 
S.S.R.O. acknowledges funding from NSF Career grant AST-1650486.
J.M.D.K. gratefully acknowledges funding from the German Research Foundation (DFG) in the form of an Emmy Noether Research Group (grant number KR4801/1-1) and a DFG Sachbeihilfe Grant (grant number KR4801/2-1), from the European Research Council (ERC) under the European Union's Horizon 2020 research and innovation programme via the ERC Starting Grant MUSTANG (grant agreement number 714907), and from Sonderforschungsbereich SFB 881 ``The Milky Way System'' (subproject B2) of the DFG. 
JBP acknowledges UNAM-DGAPA-PAPIIT support through grant number IN-111-219. 

\end{acknowledgement}

\bibliographystyle{spbasic}
\bibliography{bib_ch8}
\end{document}